\documentclass[sigconf, nonacm]{acmart}

\usepackage{enumitem}
\usepackage[T1]{fontenc}

\usepackage{textcomp}
\usepackage{xcolor}
\usepackage{listings} 
\usepackage{graphicx,graphics}
\usepackage{subfigure}
\usepackage{balance}  
\usepackage{booktabs} 
\usepackage{array}
\usepackage{mathrsfs}
\usepackage{tabularx}
\usepackage{bm}
\usepackage{multirow}
\usepackage{amsmath,amssymb}
\usepackage[linesnumbered,ruled,vlined]{algorithm2e}

\usepackage{caption}
\usepackage{url}
\usepackage{makecell}
\usepackage{booktabs}
\usepackage{listings}
\usepackage{graphicx,wrapfig,lipsum}
\usepackage{threeparttable}

\theoremstyle{definition}
\newtheorem{definition}{Definition}[section]

\theoremstyle{remark}

\SetKwInput{KwInput}{Input}                
\SetKwInput{KwOutput}{Output}              

\newtheorem{example}{Example}

\newcommand\vldbdoi{XX.XX/XXX.XX}
\newcommand\vldbpages{XXX-XXX}
\newcommand\vldbvolume{16}
\newcommand\vldbissue{1}
\newcommand\vldbyear{2023}
\newcommand\vldbauthors{\authors}
\newcommand\vldbtitle{\shorttitle} 
\newcommand\vldbavailabilityurl{URL_TO_YOUR_ARTIFACTS}
\newcommand\vldbpagestyle{plain}

\newcommand{\Method}[0]{{\fontfamily{cmss}\selectfont 
AWM}}

\newcommand{\DataPreprocessori}[0]{Data Collection \& Preprocessing Module}
\newcommand{\DataPreprocess}[0]{DCPM}

\newcommand{\Offlineori}[0]{Offline Training Module}
\newcommand{\Offline}[0]{OTM}

\newcommand{\Onlineori}[0]{Online Workload Mining Module}
\newcommand{\Online}[0]{OWMM}

\newcommand{\Optimizeori}[0]{Pattern-based Optimizing Module}
\newcommand{\Optimize}[0]{POM}

\newcommand{\DataCollect}[0]{Data Collection Layer}
\newcommand{\BERT}[0]{Workload Semantic Embedding Layer}
\newcommand{\BERTEmb}[0]{FM-Server}
\newcommand{\XGBInput}[0]{Execution Feature Process Layer}

\newcommand{\XGB}[0]{Workload Classifying Layer}
\newcommand{\Oracle}[0]{Markov-based Pattern Mining Layer}

\newcommand{\Label}[0]{Automatic Label Collection}
\newcommand{\Topo}[0]{Dependency aware Multi-query Optimizer}

\newcommand{\CFLines}[3]{ (cf. Lines #1-#2 of Algorithm~\ref{alg:#3})}

\newcommand{\MRED}[1]{#1}

\newcommand{\up}{\vspace*{-0.1in}}

\newcommand{\alione}{AQL-N}
\newcommand{\alifour}{AQL-L}
\newcommand{\opendata}{OSQL}

\newcommand{\pacc}{Pattern Precision}
\newcommand{\pnum}{\# of pattern}
\newcommand{\claacc}{F1}
\newcommand{\clatime}{Time}
\newcommand{\latency}{Latency}

\begin{document}
\title{Real-time Workload Pattern Analysis for Large-scale Cloud Databases}







\author{
Jiaqi Wang$^{1, 4  }$,
Tianyi Li$^{2}$,
Anni Wang$^{3}$,
Xiaoze Liu$^{1, 4  }$, 
Lu Chen$^{1 }$,  
Jie Chen$^{3 }$,
Jianye Liu$^{3 }$,
Junyang Wu$^{1, 4}$, 
Feifei Li$^{3 }$, 
Yunjun Gao$^{1 }$}
\affiliation{%
  \institution{
  {\large$^{1}$}College of Computer Science, Zhejiang University, Hangzhou, China \\
  {\large$^{2}$}Aalborg University, Aalborg, Denmark  $\;$
  {\large$^{3}$}Alibaba Group, Hangzhou, China\\
  {\large$^{4}$}Alibaba-Zhejiang University Joint Institute of Frontier Technologies, Hangzhou, China \\
{\large$^{1 }$}\{jqiwang, xiaoze, luchen, wujunyang,  gaoyj\}@zju.edu.cn  \hspace{0.01pt} {\large$^{2}$}tianyi@cs.aau.dk
\\ {\large$^{3 }$}\{wanganni.wan, aiao.cj,  lifeifei\}@alibaba-inc.com, {\large$^{3 }$}jianye.liu@antfin.com
  }
}

\begin{abstract}
Hosting database services on cloud systems has become a common practice. This has led to the increasing volume of database workloads, which provides the opportunity for pattern analysis.
Discovering workload patterns from a business logic perspective is conducive to better understanding the trends and characteristics of the database system. However, existing workload pattern discovery systems are not suitable for large-scale cloud databases which are commonly employed by the industry. This is because the workload patterns of large-scale cloud databases are generally far more complicated than those of ordinary databases.

In this paper, we propose Alibaba Workload Miner (\Method{}), a real-time system for discovering workload patterns in complicated large-scale workloads.
\Method{} encodes and discovers the SQL query patterns logged from user requests and optimizes the querying processing based on the discovered patterns. 
First, \DataPreprocessori{} collects streaming query logs and encodes them into high-dimensional feature embeddings with rich semantic contexts and execution features.
Next, \Onlineori{} separates encoded query by business groups and  discovers the workload patterns for each group.
Meanwhile, \Offlineori{} collects labels and trains the classification model using the labels. 
Finally, \Optimizeori{}
optimizes query processing in cloud databases by exploiting discovered patterns.
Extensive experimental results on one synthetic dataset and two real-life datasets (extracted from Alibaba Cloud databases) show that \Method{} enhances the accuracy of pattern discovery by 66\% and reduce the latency of online inference by 22\%, compared with the state-of-the-arts.
\end{abstract}

\maketitle

\pagestyle{\vldbpagestyle}
\begingroup\small\noindent\raggedright\textbf{PVLDB Reference Format:}\\
\vldbauthors. \vldbtitle. PVLDB, \vldbvolume(\vldbissue): \vldbpages, \vldbyear.\\
\href{https://doi.org/\vldbdoi}{doi:\vldbdoi}
\endgroup
\begingroup
\renewcommand\thefootnote{}\footnote{\noindent
This work is licensed under the Creative Commons BY-NC-ND 4.0 International License. Visit \url{https://creativecommons.org/licenses/by-nc-nd/4.0/} to view a copy of this license. For any use beyond those covered by this license, obtain permission by emailing \href{mailto:info@vldb.org}{info@vldb.org}. Copyright is held by the owner/author(s). Publication rights licensed to the VLDB Endowment. \\
\raggedright Proceedings of the VLDB Endowment, Vol. \vldbvolume, No. \vldbissue\ %
ISSN 2150-8097. \\
\href{https://doi.org/\vldbdoi}{doi:\vldbdoi} \\
}\addtocounter{footnote}{-1}\endgroup


\begin{figure}[t]
    \includegraphics[width=.45\textwidth]{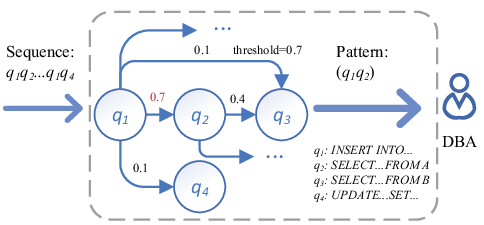}
    \up
    \caption{An example of workload pattern discovery.}
    \label{fig:mining_demo}
     \up
     \vspace{-1mm}
    \end{figure}

\section{Introduction}

\label{sec:intro}



Enterprises and consumers are increasingly hosting their services on cloud database systems, e.g., Alibaba Cloud Relational Database Service (RDS)~\cite{alibabaclouddatabase}, AWS RDS~\cite{amazon2015amazon}, Microsoft Azure SQL Database~\cite{copeland2015microsoft}, and Google Cloud SQL~\cite{krishnan2015building}.   
With the advancement of cloud database technology, these systems can now support a wider range of optimizations, e.g., automatic query re-write~\cite{SIGMOD19Index}, root cause diagnose~\cite{yoon2016dbsherlock,ma2020diagnosing, PinSQL}, automatic scale-up instances~\cite{PinSQL}, and automatic tuning~\cite{VLDB09iTuned}. The optimization techniques~\cite{SIGMOD19Index,yoon2016dbsherlock,ma2020diagnosing, PinSQL, VLDB09iTuned} have significantly improved the performance of Database Management System (DBMS), in terms of both the overall system and the specific query execution. Workload pattern discovery can be used to further optimize DBMS performance by providing statistics about entire patterns in the workload, which can significantly enrich the context of existing reports~\cite{tran2015oracle}. An example of workload pattern mining is shown as follows.

\vspace{1mm}
\begin{example}~\label{ex1}
Fig.~\ref{fig:mining_demo} shows an example of workload pattern discovery. Given a sequence of queries \textit{$q_1q_2...q_1q_4$} executed by a DBMS and a probability threshold $0.7$, only \textit{$q_1q_2$} is returned as its probability of occurrence is no smaller than 0.7. On the other hand, \textit{$q_1q_2q_3$} is not returned because its probability of occurrence 0.28 ($=$ 0.7$\times$0.4) is smaller than 0.7.
\vspace{-1mm}
\end{example}

Example~\ref{ex1} shows a frequently executed code sequence \textit{$q_1q_2$}. Such information is crucial for database administrators (DBAs) and application developers to optimize quering processing and improve application performance. Note that, the frequency of executed code paths may be unavailable without pattern mining, especially if the components of applications work as black boxes~\cite{tran2015oracle}. Workload pattern mining is thus essential in inferring the business logic and user activities. For example, the DBAs for database-as-a-service (DaaS) offerings (e.g., Oracle Cloud and Alibaba Cloud) require knowledge about the business logic. However, due to permission issue, the DBAs generally do not have access to the source code of the applications. In this case, workload pattern mining can provide DBAs with a useful model for tracing workloads. The model offers the DBAs better insight into the applications running in the database and provides developers of applications with the working flow of the application's logic~\cite{tran2015oracle}.

However, analyzing workload patterns from large-scale industrial cloud databases is challenging, because the query log often contains queries from multiple business logics rather than a single one.  To the best of our knowledge, WI~\cite{tran2015oracle}, the only existing study for workload pattern analysis on databases, cannot distinguish the mixed queries and thus cannot be applied to industrial applications.

\begin{figure}[t]
    \includegraphics[width=.47\textwidth]{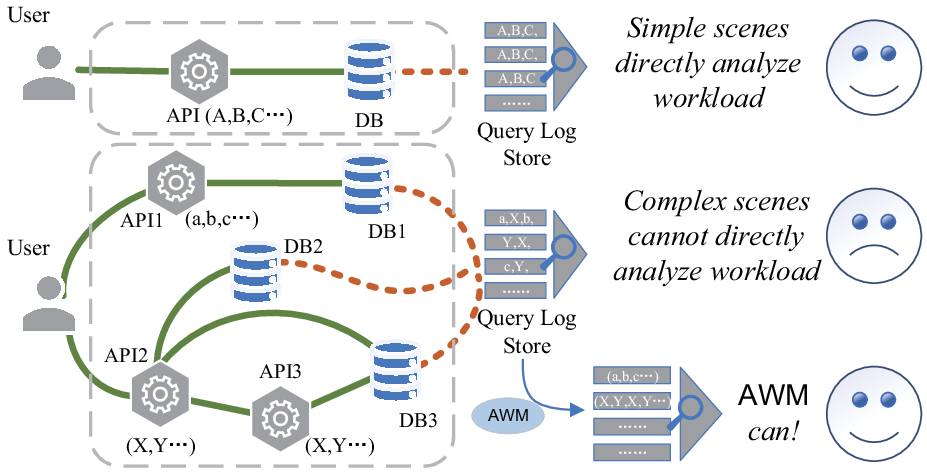}
    \up
    \caption{Examples of analyzing workloads from single business logic and multiple business logic.}
    \label{fig:intro_demo}
    \up\up
    \end{figure}

\vspace{-2mm}
\begin{example}\label{ex2}
Fig.~\ref{fig:intro_demo} depicts two cases of workload pattern discovery. 
In the upper case, 
queries (e.g., $A, B, C, \cdots$) from the same business logic (i.e., API) are transferred to a query log store for workload pattern discovery.  Since only a single business logic exists in this case, WI~\cite{tran2015oracle} can be employed without the need for query classification by business logic. In the lower case, queries (e.g., $A, B, C, \cdots$) from multiple business logics (i.e., API1, API2, API3) are mixedly loaded into a query log store. This results in interleaved queries from different business logics in a sequence (e.g., $\{a, X,b\}$, $\{Y, X\}$, and $\{c, Y\}$), which serves as the input of the subsequent workload pattern discovery.
This scenario is common in industry applications. WI~\cite{tran2015oracle}, however, cannot distinguish queries from multiple business logics and thus cannot discover workload patterns correctly. The lower case highlights the necessity of distinguishing patterns from multiple business logics in industrial applications.
\end{example}
\vspace{-2mm}


In this paper, our goal is to develop a system for discovering workload patterns from large-scale query logs and utilizing the discovered patterns to optimize subsequent query processing. Specifically, the system needs to be capable of (i) classifying queries by types of business logic while maintaining users' privacy, (ii) efficiently performing pattern discovery on each business logic, and (iii) leveraging the discovered patterns to optimize query processing in the cloud database. However, there are four key issues that must be addressed to achieve these objectives.

\begin{itemize}[topsep=0pt,itemsep=0pt,parsep=0pt,partopsep=0pt,leftmargin=*]
    \item \textbf{\textit{Scalability.} }
    The system should be designed to handle large-scale datasets, enabling it to train classifiers and identify workload patterns across billions of queries. 
    \item \textbf{\textit{Privacy.} }
    A potential classification method is to analyze the App ID and code, which can serve as labels for identifying the business logic. However, such information is considered private and thus is not accessible~\cite{PinSQL} to the system. Therefore, an alternative approach is needed that can classify the data with limited labels. 
 
    \item \textbf{\textit{Accuracy.} } Achieving high classification accuracy
can be challenging when the true label of a query is mostly unavailable due to user privacy concerns.  Thus, it is important to develop classification algorithms that can achieve high accuracy even with limited true labels.
 
   \item \textbf{\textit{Optimization.}} Discovering patterns in query workloads can provide valuable insights into optimizing future query processing in cloud DBMS. 
   However, existing pattern mining systems typically only provide pattern mining results without clear optimization guidelines.  This makes it difficult for users who lack knowledge of the underlying logic of database engines to design their own optimization strategies. We thus aim to not only perform pattern discovery but also offer clear optimization guidelines based on discovering results through a user interface.
   
\end{itemize}

We present \underline{A}libaba \underline{W}orkload \underline{M}iner (\Method{}), a comprehensive system that consists of four key modules: \DataPreprocessori{} (\DataPreprocess{}), \Offlineori{} (\Offline{}), \Onlineori{} (\Online{}), and \Optimizeori{} (\Optimize{}).
The \DataPreprocess{} module is responsible for large-scale data collection and preprocessing. It consists of two feature extraction layers: (i) \BERT{}, which utilizes pre-trained foundation models~\cite{bert} to extract latent information from SQL queries and provide a rich source of information for classification; and (ii) \XGBInput{}, which encodes execution features, such as response time, into a unified feature that serves as input for the classifier model.

To achieve high \textit{scalability},
\Offline{} handles heavy workloads efficiently (e.g., label collection and model training), which enables \Online{} to infer the trained model and obtain workload patterns in real-time.
To preserve \textit{privacy}, \Offline{} incorporates a novel automatic label method. This method allows users to flexibly select the labels that can be shared as training data. To achieve high \textit{accuracy}, \Online{} employs an effective classifier that categorizes patterns by business logics, with the use of rich features provided by \DataPreprocess{}. Continuing Example~\ref{ex2}, \Method{}' classifier distinguishes $X$ from $a, b$ such that queries from distinct business logic are separated, e.g., $a,b,c,\cdots$ and $X, Y, X, Y \cdots$. This enables effective discovery of workload patterns.
To achieve \textit{optimization}, \Optimize{} provides users with clear guidelines for optimizing query processing. This allows users to define the business logic-related dependencies for SQL queries by themselves, even if they are not experts in DBMS. In summary, the paper makes the following contributions:

\begin{itemize}[topsep=0pt,itemsep=0pt,parsep=0pt,partopsep=0pt,leftmargin=*]
\item{} We develop \Method{}, an autonomous workload pattern mining system for cloud databases, which discovers frequent workload patterns and provides guidelines for optimizing query processing. To the best of our knowledge, \Method{} is the first system that has been successfully deployed on large-scale cloud databases.

\item{} We present a high-efficient \DataPreprocess{}. This module can not only process 
large-scale data in real-time but also extract sufficient latent information from pre-trained foundation models.

\item{} We develop a two-stage framework, \Offline{} and \Online{}, for autonomous workload pattern mining.  
\Offline{} handles complicated tasks with heavy workloads; while \Online{} infers the trained model to obtain the discovered patterns promptly in real-time.

\item{} We propose an optimization scheme in \Optimize{}. It analyzes the discovered patterns by exploiting dependency graphs and provides users with clear guidelines through a user interface. With the guidelines, users can optimize query processing without requiring professional knowledge.

\item{} We conduct extensive experiments on one synthetic dataset and two real-life datasets. The experimental results suggest that compared with the state-of-the-arts, \Method{} achieves a 70\% improvement in the accuracy of pattern discovery and a 25\% reduction in the latency of online inference.

\end{itemize}

The rest of the paper is organized as follows. 
We present preliminaries in Section~\ref{sec:preliminaries} and give an overview of the proposed system
in Section~\ref{sec:overview}. Section~\ref{sec:preprocessing} details the \DataPreprocess{}, and Section~\ref{sec:online} covers the \Online{}. Sections~\ref{sec:offline} and~\ref{sec:optimize} present \Offline{} and \Optimize{}, respectively. Section~\ref{sec:experiment} reports the experimental results. Section~\ref{sec:related work} reviews the related work. Section~\ref{sec:conclusions} concludes and offers
directions for future work.



\section{Preliminaries}
\label{sec:preliminaries}

\subsection{Markov chain}

A Markov chain is a type of stochastic model describing a sequence of possible events. A first-order Markov chain considers that the future state depends only on the current state and has nothing to do with the previous state. Put differently, first-order Markov models are memoryless. The $m_{th}$-order Markov chain is an extension of the first-order Markov chain, which considers that the future state depends on the past $n$ states. $m_{th}$-order Markov chains satisfy the following equation of conditional probability:
\begin{equation}
\begin{split}
   &P\left(X_t=x_t \mid X_{t-1}=x_{t-1}, \ldots, X_1=x_1\right) \\
    &=P\left(X_t=x_t \mid X_{t-1}=x_{t-1}, \ldots, X_{t-m}=x_{t-m}\right) \\
\end{split}
\end{equation}

\noindent where $X_i$ and $x_i$ represents the $i_{th}$ state of the random variable $X$ and its value, and $t$ is the future state to be identified.
The change in the state of a random variable over time steps in a Markov chain is called transition. A transition matrix is generally adopted to describe the structure of a Markov chain. It represents the properties that the Markov chain exhibits during the transition process. A state transition probability is defined as the conditional probability between random variables in a Markov chain.

Markov chains can be exploited to predict and identify the context of SQL queries in the field of workload patterns~\cite{tran2015oracle}. 
Considering each SQL query as a state of the random variable, Markov chains formulate the transition probabilities among queries in SQL context. Since Markov chains has proven effective for workload pattern mining~\cite{tran2015oracle}, we incorporate Markov chains into \Method{}.

\subsection{Minimum Description Length principle}

The Minimum Description Length (MDL) principle~\cite{mdl2007} is a commonly used model selection principle. MDL is particularly suitable for dealing with the selection, prediction, and estimation of complicated models. 
The quantity of interest (a model or/and parameters) is called a hypothesis. The best hypothesis defined by MDL describes the regularities of data, such that employing it on data compression achieves the highest compression ratio~\cite{li2020compression,li2021trace}.
MDL can be formulated according to the maximum a posteriori (MAP) estimation~\cite{mdl2007}:
$L(D)=\min _{h \in \mathcal{H}}(L(D \mid h)+L(h))$, where $L$ denotes description length. The first term in the formula is the description length of the training data $D$ given the hypothesis $h$. The second term in the formula is the description length of $h$ in the hypothesis space $\mathcal{H}$. 
The MDL principle selects the hypothesis $h$ that minimizes the sum of these two description lengths. MDL is essentially a balance between model complexity and the number of errors. More specficially, it aims to choose a shorter hypothesis that generates fewer errors. 

\begin{figure*}[t]
\vspace{-8mm}
    \includegraphics[width=.98\textwidth]{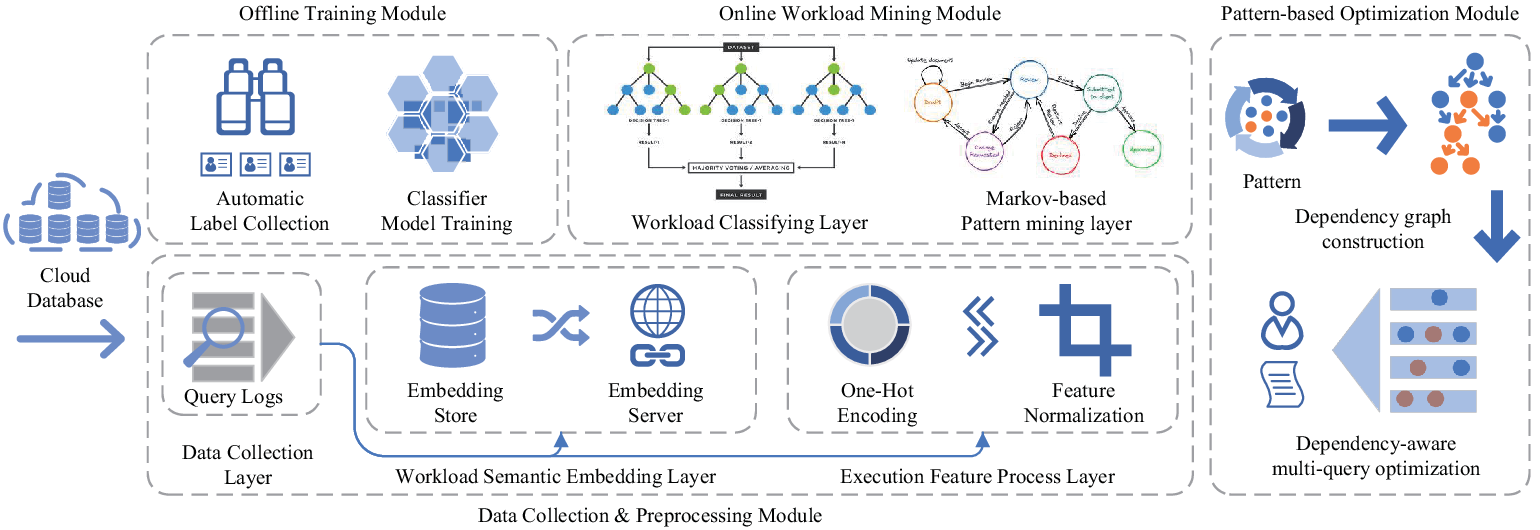}
    \up
    \caption{System overview}
    \label{fig:introduction} 
    \up\up
    \vspace{3mm}
    \end{figure*}

\subsection{Problem Definition}
\label{sec:problem}

\begin{definition}[\textbf{SQL template}]
    An SQL template (or SQL digest) is a composite of multiple queries that are structurally similar but may have different literal values.  
   An SQL template replaces hard-coded values in the statement with a placeholder (e.g '?').
    \end{definition} 
\begin{example}
   An SQL template "SELECT  *  FROM item\_table
     WHERE item\_id = ?" might include the following children queries:
     \begin{itemize}[topsep=0pt,itemsep=0pt,parsep=0pt,partopsep=0pt,leftmargin=*]
        \item SELECT  *  FROM \;item\_table \;WHERE item\_id = ABCDEF
        \item SELECT  *  FROM \;item\_table \;WHERE item\_id = GHIJKFM
        \item SELECT  *  FROM \;item\_table \;WHERE item\_id = NOPQRS
    \end{itemize}   
\end{example}

Using templates is sufficient for most optimization techniques in practice because the absence of special values has minimal impact on request behavior. A workload pattern is represented by SQL templates, which are interpretable sequences of SQL context in the database's request behavior. These templates characterize workloads and unveil query patterns that correspond to applications. 
We formally define it as follows.
\begin{definition}[\textbf{workload pattern}] Given a threshold $\alpha$,
a workload pattern $p$ is a sequence of queries $q_1, q_2, \cdots, q_m$, whose occurrence probability is no smaller than $\alpha$.
\end{definition}

\begin{definition}[\textbf{workload pattern discovery}]
    Given a query log $Q$ containing a set of queries $q_1, q_2, \cdots, q_n$, workload pattern discovery aims to find a set of workload patterns $P=\{p_i|i=1,2,\cdots,n\}$.  

\end{definition}



\section{System Overview}
\label{sec:overview}

\begin{table}[t]\small
    \caption{Execution feature collected in the query log store}
    \label{tab:feature_list}
    \vspace*{-4mm}
    \resizebox{\linewidth}{!}{\begin{tabular}{lll}
        \toprule
        Feature              &  & Description                                                \\ \cline{1-1} \cline{3-3} 
        \textit{lock\_wait\_time}     &  & Waiting time to access data  \\
        \textit{logical\_read}        &  &  The number of blocks read from memory       \\
        \textit{rows\_examined}       &  & The number of rows scanned\\
        \textit{rows\_returned}       &  & The number of rows of data returned                        \\
        \textit{rows\_updated}        &  & The number of updated data rows                            \\
        \textit{rt}                   &  & The response time of the transaction                       \\
        \textit{timestamp}            &  & The time that the query begins to execute            \\
        \textit{physical\_sync\_read} &  & The number of blocks read from disk           \\
        \textit{database}             &  & Name of the database stated in the query                   \\
        \textit{error\_code}          &  & Execution error code                                       \\
        \textit{origin\_host}         &  & Source database address                                    \\
        \textit{sql\_type }           &  & SQL statement type (e.g., INSERT)                          \\
        \textit{sql}                  &  & SQL text\\                   \bottomrule   
        \end{tabular}
    }
    \vspace*{-3mm}
\end{table}

We develop a workload pattern discovery system \Method{}, which actively collects the SQL query patterns logged from the user requests, and then analyzes and provides strategies for optimizing those patterns.  
We have deployed \Method{} on Alibaba Cloud Database Autonomous Service (DAS)~\footnote{https://www.alibabacloud.com/product/das}. The system encompasses four modules: \DataPreprocessori{} (\DataPreprocess{}), \Offlineori{} (\Offline{}), \Onlineori{} (\Online{}), and \Optimizeori{} (\Optimize{}). Fig.~\ref{fig:introduction} gives an overview of \Method{}.

\begin{figure}[t]
\includegraphics[width=.47\textwidth]{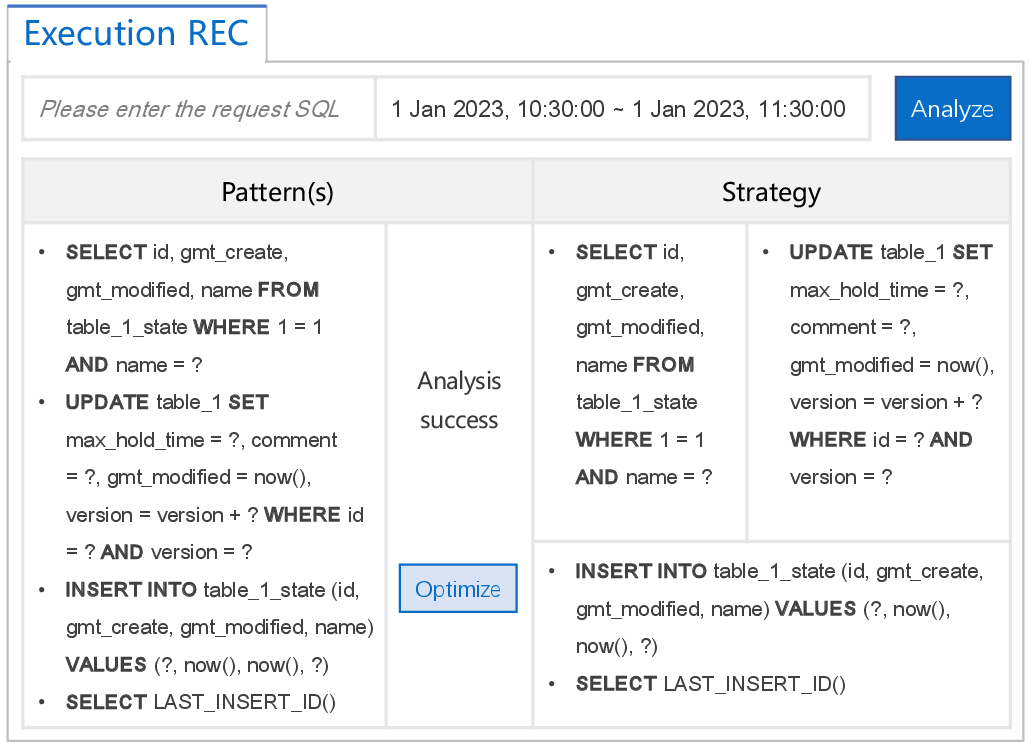}
\vspace{-2mm}
\caption{Demonstration of the user interface.}
\label{fig:interface}
\vspace{-4mm}
\end{figure}

\DataPreprocess{} processes data with three layers: \DataCollect{}, \BERT{} and \XGBInput{}. \DataCollect{}
collects and pre-processes the streaming raw data (Performance Metrics data and Query Logs data) from millions of database instances in real-time and stores the processed data in a local storage. 
\BERT{} encodes the queries into high-dimensional feature embeddings with rich semantic contexts. \XGBInput{} processes the execution features into uniformed embeddings.

\Offline{} automatically collects training labels 
and trains classifiers for \XGB{} with the labels and the pre-processed feature vectors. \Online{} discovers workload patterns in a fully online fashion. It has two layers: \XGB{} and \Oracle{}.
\XGB{} encompasses an effective classifier that is trained offline by \Offline{} with few labels. It categorizes query logs by business groups in real-time.
\Oracle{} receives the classified query logs and performs pattern discovery on them.

\Optimize{} offers code optimization strategies presented in a user interface to cloud database users. 
It automatically idnetifies potential optimizations in a set of SQL queries that may arise from sub-optimal business logic code. \Optimize{} allows users to flexibly define the business logic-related dependencies for SQL queries.  

We develop a user interface (UI) that provides a visualization of identified patterns and optimization strategies for users based on their requests. We integrate UI into Alibaba Cloud Database Autonomy Service System, part of which is shown in Fig.~\ref{fig:interface}. In particular, when a user submits a SQL query, \Method{} analyzes it and displays corresponding patterns and optimization strategies in the UI. Among the strategies shown, SQL queries that appear in the parallel cells of a row can be parallelized.



\section{\DataPreprocessori{}}
\label{sec:preprocessing}

\DataPreprocess{} contains three layers: \DataCollect{}, \BERT{}, and \XGBInput{}. When the DBMS executes a SQL query, \DataCollect{} firstly collects all the information related to the query, and then \BERT{} encodes the SQL queries into semantic features for a unified process. Meanwhile, \XGBInput{} is applied to obtain the vectorized execution features of the queries. Finally, the two features are fed into both \Online{} and \Offline{} as an integrated feature of each query.

\subsection{Data Collection Layer}
\DataCollect{} collects large-scale streaming query logs data via the Audit log of DB engines~\cite{PinSQL}.
The query log contains two kinds of information for query execution: the raw query text and the recorded execution data (e.g., query response time).
We provide the collected features and the corresponding description in Table~\ref{tab:feature_list}. Since the data is asynchronously loaded into Alibaba Cloud LogStore~\cite{cao2021logstore} in real-time, it has little impact on database instances~\cite{mozafari2013performance, theriault1998oracle}. Note that, we delete the collected data every three days to avoid storage overflow. 
We will detail how we encode the input query into a unified feature in Section~\ref{sec:bert} and Section~\ref{sec:feature}. 
 

\subsection{\BERT{}}
\label{sec:bert}

\begin{algorithm}[tb]
    \small
    \DontPrintSemicolon
      \SetKwProg{Fn}{def}{:}{}
      \KwInput{The input queries $Q$, batch size $b$,  pooling method $p$, and the embedding store $D$}
      \KwOutput{Output vector $Z_o \in \mathbb{R}^{|Q|\times d}$ }
        \SetKwFunction{FBERTEmb}{\BERTEmb{}}
        \SetKwFunction{FEmbStore}{Embedding\_Store}
        \SetKwFunction{FTokenizer}{Tokenizer}
        \SetKwFunction{FLanguageModel}{Language\_Model}
        \SetKwFunction{FMask}{Pooling\_Mask}

        $\text{idx\_new}$ $\leftarrow []$\tcp*{id of queries to be embedded}
        $Q_n$ $\leftarrow []$\;
        $Z_o \leftarrow 0 ^{|Q|\times d}$\;
        \For{$i, q \in \textsf{enumerate}(Q)$}{

            \If{$q \not\in D$}{
                $\text{idx\_new}$.append($i$)\tcp*{add an id to be embedded}
                $Q_n$.append($q$)\;
            }
            \Else(){
                $Z_o[i]\leftarrow D[q]$\tcp*{extract value directly from D} 
            }
        }
        $Z_n \leftarrow $\FBERTEmb{$Q_n$, $b$, $p$}\;
        $Z_o[\text{idx\_new}]\leftarrow Z_n$ \;
        \For(){$i, q \in enumerate(Q_n)$}{
            $D[q] \leftarrow  Z_n[i]$\;
        }
        \KwRet{$Z_o$}\;
    \caption{\BERT{}}
    \label{alg:bert_layer}
\end{algorithm}
\BERT{} pre-processes data by providing sufficient features from the SQL text corpus in a unified vector. This enables \XGB{} to achieve higher performance with limited training labels. 


\subsubsection{Deploying the Embedding Server of Foundation Models}
\BERT{} serves an API online with GPU acceleration to map the SQL query text into a high dimensional embedding space, namely \BERTEmb{}. Here, we use the pre-trained foundation models (FMs)~\cite{bert,liu2019roberta, transformers} to embed the SQL text. 
This is because FMs can learn latent information from the web, which enhances the classification with fewer training labels. 
The pre-trained FMs have proven effective for subsequent tasks such as knowledge graphs~\cite{LargeEA22} and data integration~\cite{ge2021collaborem}. Since SQL query text is generally text data generated with meaningful statements, incorporating FMs can result in better accuracy~\cite{PreQR}.
To implement pre-trained FMs, we adopt the transformers\footnote{https://github.com/huggingface/transformers/} library from hugging face for obtaining an accurate and reliable service. Here, 
the FM maps an input sequence of SQL text token representations $(x_1, ..., x_n)$ to a sequence of continuous representations $\mathbf{z} = (z_1, ..., z_n)$, where $x$ is a token, $(x_1, ..., x_n)$ is the whole query text, and $z_i \in \mathbb{R}^d$ is the vector representation of dimension $d$ indicating the embedding of $x_i$. An example is shown as follows. 

\begin{example}
\label{example:bert_input}
Given an SQL query text "SELECT id, name FROM user\_table WHERE id=5", and a simple tokenizer dividing the text by space,
we get a list of tokens $T=$ ["SELECT", "id,", "name", "FROM", "user\_table", "WHERE", "id=5"]. For this SQL text, the FM outputs a vector $Z \in \mathbb{R}^{|T|\times d}$ where $|T|$ indicates the length of the token list. 
\end{example}

In practice, more complex tokenizers~\cite{bert} are employed in the upstream tasks of pre-trained FMs. Here, we process the queries by batch to speed up the overall procedure. Thus, with a batch of query texts, a vector $Z_b \in \mathbb{R}^{b\times max(|T|) \times d}$ is obtained, where $b$ is the batch size. Note that, the middle dimension of $Z_b$ is $max(|T|)$, which indicates the FM pads the output in order to align the output length of all sequences. Suppose we have a batch, in which the shortest sequence length is $t$, then the output of this sequence will be a vector of $max(|T|) \times d$. In this vector, only the prefix of $t\times d$ has the valid semantic value, and the rest is padded with a constant number zero by default.

After applying FM, we map the sequence of vectors  $Z_b$ for each SQL statement into a unified size vector $Z_o \in \mathbb{R}^{b\times d}$, to be prepared for the downstream tasks. This incurs the demand for pooling, which reduces the sequence of vectors into one vector. We support two types of pooling methods: (1) max pooling, which returns the maximum value along the given dimension; and (2) mean pooling, which returns the averaged value along the given dimension. In each pooling method, we pool the result matrix along the sequence dimension. Thus, for each value of the output vector, the result is reduced from all the tokens acquired by FM, i.e., the original $Z_b \in \mathbb{R}^{b\times max(|T|) \times d}$ is transformed into  $Z_o \in \mathbb{R}^{b\times d}$ by pooling.

Since the output is a padded vector, the padded value will have an adverse effect on the result. By using the default constant, we discuss the following two cases: (i) when using max pooling, we may be unable to get a negative value from the FM; and (ii) when using mean pooling, we may obtain a relatively smaller output vector in terms of the norm. To avoid these, we first augment the output (e.g., for max pooling, we pad the output vectors with $-\infty$), and then calculate the mean or max value.

\begin{algorithm}[tb]
    \small
    \DontPrintSemicolon
      \SetKwProg{Fn}{def}{:}{}
      \KwInput{The input queries $Q$, batch size $b$,  pooling method $p$}
      \KwOutput{Output vector $Z$ }
        
        \SetKwFunction{FBERTEmb}{\BERTEmb{}}
        \SetKwFunction{FEmbStore}{Embedding\_Store}
        \SetKwFunction{FTokenizer}{Tokenizer}
        \SetKwFunction{FLanguageModel}{Language\_Model}
        \SetKwFunction{FMask}{Pooling\_Mask}

        \Fn{\FBERTEmb{$Q$, $b$, $p$}}{

            $\textit{batch\_idx} \leftarrow$  0\; 

            $\textit{result}$ $\leftarrow []$

            \While{$\textit{batch\_idx} < |Q|$}{
                $\textit{batch\_end} = min(\textit{batch\_idx} +b, |Q|)$\;
                $query\_batch = Q[\textit{batch\_idx}: \textit{batch\_end}]$\;
                $T_b \leftarrow$ \FTokenizer{$query\_batch$}\;
                $S_T \leftarrow len(T \; \text{for} \; T \; \text{in} T_b)$\;
                $Z_b \leftarrow $\FMask{ \FLanguageModel{$T_b$}, $S_T$, $p$}\;

                \If{$p$ \text{is "max\_pooling"}}{

                    $Z\leftarrow Z_b.\text{max(axis=1,keepdim=False)}$\;
                }
                \If{$p$ \text{is "mean\_pooling"}}{

                    $Z\leftarrow Z_b.\text{mean(axis=1,keepdim=False)}$ \; 
                }

                $\textit{batch\_idx}\leftarrow \textit{batch\_idx} + b$\;
                $\textit{result}\text{.append(}Z\text{)}$\;
            }
            \KwRet{$\text{concatent(result)}$}
        }
    \caption{\BERTEmb{}}
    \label{alg:bert_emb}
\end{algorithm}

\subsubsection{Embedding Store}
\label{subsubsec:embedding_store}
We observe that most of the SQL queries executed are repeated. Thus, we develop an embedding store to cache the output feature embeddings. 
The embedding store should satisfy the following two requirements.
First, it should follow idempotence~\cite{richardson2008restful} of \BERT{}. This means that, for one specific input, \BERT{} can be applied several times, but the resulting state of one call should be indistinguishable from the consequent state of multiple calls. This results in a consistent handling function of duplicate requests received by the API. Duplicate requests may arrive unintentionally or intentionally. For example, a user may send several duplicate queries due to timeout or network issues. However, the output embeddings of FMs may be different in terms of the batch context, which is not desirable.
Second, \BERTEmb{} requires massive computational resources. Thus, applying it to infer massive queries in real-time is unrealistic and inefficient (cf. Section~\ref{sec:experiment}).

To address the above-mentioned issue, we hash the SQL text of each input and store its unique embedding index by a global hash map. 
For each newly emerged query, we first check if the query text has been computed. If it has, we directly output the stored embedding; if it has not, we compute the feature embedding with \BERTEmb{}, and then store it in the embedding store.

The pseudo-code of \BERT{} is shown in Algorithm.~\ref{alg:bert_layer}. Given the input $Q$, which is the set of SQL query texts, the algorithm first initializes the output memory space and other supporting variables \CFLines{1}{3}{bert_layer}. Then, for each query, we check if its embedding has already been calculated. If it has, the value is directly obtained~\CFLines{4}{9}{bert_layer}. The embedding queries that are not cached by the embedding store are then calculated online and merged into all features~\CFLines{10}{11}{bert_layer}. Next, we update the embedding store and return the currently requested vectors. 

The implementation of  \BERTEmb{} is shown in Algorithm~\ref{alg:bert_emb}, which follows a batched design to infer the foundation model. For each batch, we first compute the tokens and other supplementary features~\CFLines{4}{8}{bert_emb}. Then, we apply FM to the queries and assign corresponding values to padded positions for different pooling methods. To pad the right value for reduction, we calculate a mask for each query token sequence, indicating the position of the padded value. If the pooling method is "max pooling", a very small number ($-\infty$) is padded. For "mean pooling", we first pad the value to 0 and then scale the whole batch~\CFLines{9}{13}{bert_emb}.
Finally, we reduce the output size by pooling the embedding and proceed to the next batch~\CFLines{10}{15}{bert_emb}.



\vspace{-3mm}
\subsection{\XGBInput{}}
\label{sec:feature}

The classification model generally can only take numerical value as the input to make the regression prediction. In this section,
we detail how we use \XGBInput{} to preprocess the recorded execution data into numerical data, in order to sufficiently learn all feature information of each query.

The data features and the corresponding description are shown in Table \ref{tab:feature_list}. We divide all features into two primary categories: (i) features with numerical meaning (e.g., \textit{rows\_examined}, \textit{logical\_read}, and \textit{rt}) and (ii) features without numerical meaning (e.g., \textit{origin\_host}, \textit{error\_code}, and \textit{sql\_type}). 
For the first category, we normalize them with their mean value and std. For the latter category, OneHot is adopted to encode their input variables. As a common encoding method in machine learning, OneHot encoding uses N-bit 0/1 registers to encode N states.

Note that, when numerical features follow long tail distribution, directly encoding them tends to result in high data dimension. Hence, we preprocess numerical features before label encoding or OneHot encoding to modify its representation.
For example, the \textit{rows\_examined} feature is represented as ten integer values from $1$ to $10$ based on its numerical distribution. In addition, some features have special significance when their values are zero (e.g. the query is most likely a SELECT statement when its \textit{rows\_updated} is 0), and we represent such features as specified integers.

\vspace{-2mm}
\section{\Onlineori{}}
\label{sec:online}

\Online{} acts as a real-time service. It contains two layers: \XGB{} and \Oracle{}. When a query arrives, \XGB{} first receives the features of this query from \DataPreprocess{} for classification by business groups.
Then, \Oracle{} of the corresponding business group processes this query.

\subsection{\XGB{}}
In this layer, we classify the queries into different business groups. 
The input of each query contains: the semantic embedding $Z$ based on \BERT{} and the query execution information $X$ based on \XGBInput{}. More specifically, we first concatenate the two features to obtain a unified input vector of each query, i.e., $F = [Z || X]$. 
Then, we can apply a classification model to the input embeddings, i.e., $clf(F)$, where $clf$ is the pre-trained classifier. Although \XGB{} can incorporate any kind of classifier model, it cannot be specified by users because it is a general classifier for all users and is thus hard to satisfy users' localized and specialized needs. 
This paper sets XGBoost Classifier (XGBClassifier)~\cite{XGBoost16} as the default classification model due to its high scalability.


\subsection{\Oracle{}}
This layer is to discover workload patterns from SQL queries filtered by \XGB{}, where the Markov chain model and Oracle Workload Intelligence (WI)~\cite{tran2015oracle} model are employed. In the following, we introduce this layer  by detailing the transformation from SQL query texts to templates, the Markov model selection and the workload patterns mining.

\subsubsection{Transform SQL query texts into Templates}
To handle massive SQL queries, we transform SQL query texts into SQL templates according to the equivalence of the operation.
Specifically, we first gain SQL templates~\cite{mysqldigest,tran2015oracle,ma2018query,amazonuserguide} based on its definition (cf. Section~\ref{sec:problem}). 
Next, we hash the template via a unique SQL\_ID. 
Finally, we identify a set of SQL queries as equivalent among the discovered patterns if their SQL\_IDs are identical.

\subsubsection{Select the appropriate Markov model}
We build a prefix tree for the SQL sequence $S$ for analyzing and calculating the state transition probabilities using  MDL principle~\cite{tran2015oracle}.
The prefix tree is an ordered tree 
with $max\_ord+2$ layers of nodes, where $max\_ord$ is the maximum order of the Markov model. The root node is at the first level, i.e., level zero, and the leaf nodes are at the last level, i.e., level $max\_ord+1$. Each node in the prefix tree represents an ordered sequence. The edge between nodes represents a statement, and the node stores the occurrence frequency of its represented ordered sequence.

\begin{figure}[t]
    \includegraphics[width=.39\textwidth]{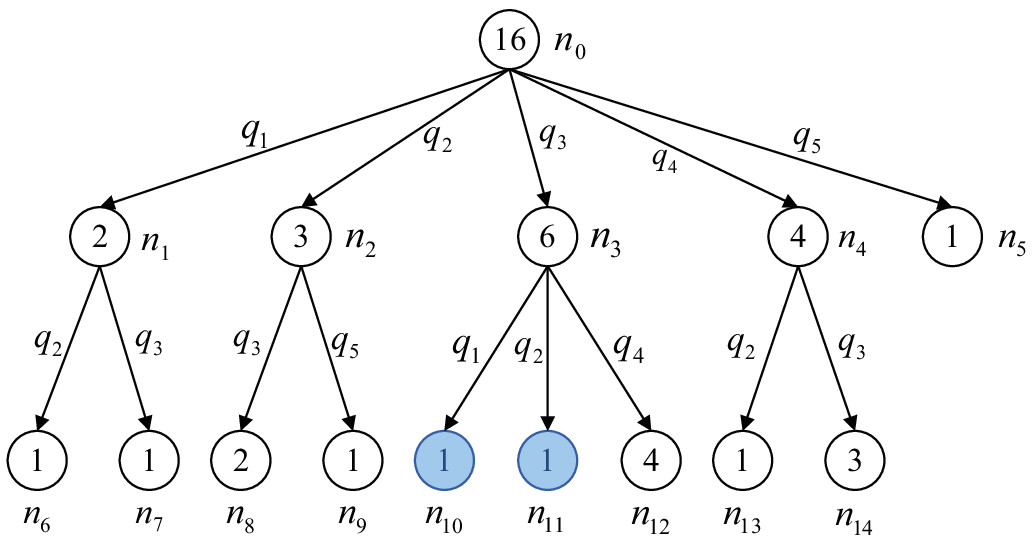}
    \up
    \caption{A prefix tree for Example~\ref{example:oracle1}.}
    \label{fig:trie}
    \up\up
    \end{figure}

\begin{example}
    \label{example:oracle1}
    Given a SQL sequence $S = q_1 q_2 q_3 q_4 q_3 q_4 q_2 q_3 q_4 q_3 q_1 q_3 \\q_4 q_3 q_2 q_5$ and the maximum order of Markov model \textit{${max\_{ord}}=1$}, a prefix tree is constructed, as shown in Fig.~\ref{fig:trie}. 
    To be specific, $n_2=3$ represents that $q_2$ appears 3 times in the sequence; $n_8=2$ represents that $q_2 q_3$ appears 2 times in the sequence.
\end{example}

In order to reduce the time cost and improve the generalization ability of the model, we prune the built prefix tree to calculate state transition probabilities before applying MDL principle.
Specifically, the state transition probabilities are computed by $\mathrm{P}\left(q_x\mid\mathrm{s}\right)=\frac{\operatorname{value}\left(n_c\right)}{\operatorname{value}\left(n_p\right)}$, 
where $s$ and $n_p$ represent the current sequence and its corresponding node, while $q_x$ and $n_c$ represent the new query and its corresponding node. However, when calculating the nodes in the same layer, the probability may be 0, which ignores the appearance possibility of some sequences and affects the subsequent model selection. Hence, it is necessary to modify the method of computing state transition probabilities according to the threshold $\tau=1 /|\operatorname{set}(S)|$:

\begin{itemize}[topsep=0pt,itemsep=0pt,parsep=0pt,partopsep=0pt,leftmargin=*]
    \item If $\mathrm{P}\left(q_x\mid\mathrm{s}\right)\geq \tau$, the transition probability follows its corresponding real distribution;
    \item If $\mathrm{P}\left(q_x\mid\mathrm{s}\right)< \tau$, the transition probability is considered to be distributed uniformly.
\end{itemize}
 
\begin{example} 
    \label{example:oracle2}
    Continuing Example~\ref{example:oracle1}
    and given  $\tau=1 /|\operatorname{set}(S)|=1/5$, 
    $P\left(q_4 \mid q_3\right)=\frac{\operatorname{value}\left(n_{12}\right) }{ \operatorname{value}\left(n_3\right)}=\frac{2}{3}>\frac{1}{5}$. Thus, $P\left(q_4 \mid q_3\right)$ is preserved as its real distribution. On the contrary, due to $P\left(q_2 \mid q_3\right)=P\left(q_1 \mid q_3\right)=\frac{\operatorname{value}\left(n_{10}\right)}  {\text { value }\left(n_3\right)}=\frac{1}{6}<\frac{1}{5}$ and $P\left(q_3 \mid q_3\right)=P\left(q_5 \mid q_3\right)=0<\frac{1}{5}$ 
    $P\left(q_2 \mid q_3\right)$, $P\left(q_1 \mid q_3\right)$, $P\left(q_3 \mid q_3\right)$ , and $P\left(q_5 \mid q_3\right)$ are all set to $\left(1-P\left(q_4 \mid q_3\right)\right) / 4=\frac{1}{12}$.
\end{example}

In order to choose the most suitable Markov model, we calculate the costs of pruned models with different orders according to MDL principle following~\cite{tran2015oracle}:

\begin{equation}
\begin{split}
    \mathbb{C}\left(S, M^{{ord }}\right)=2\times(\log { ord }+ \log m +1)\\
    +m(({ord}+1) \log |\operatorname{set}(S)|  +2 \log |S|)\\
    -\log \mathrm{P}\left(S \mid M^{{ord }}\right)
\end{split}
\end{equation}

\noindent
where $M^{ord}$ is the Markov model of ${ord}_{th}$ order; $m$ is the number of items whose probability values reach the threshold $\tau$; $\operatorname{set}(S)$ is the statement set of SQL sequence $S$; and $\mathrm{P}\left(S \mid M^{{ord }}\right)$ is the probability calculated by continuous multiplication. Continuing Example~\ref{example:oracle2} where order of Markov model is 1, $\mathrm{P}\left(S \mid M^{{ord }}\right)$ is derived as follows.
\vspace{-2mm}
\begin{equation}
\begin{split}
    \mathrm{P}\left(S \mid M^{{ord }}\right)=P\left(q_1\right) \times P\left(q_2 \mid q_1\right) \times P\left(q_3 \mid q_2\right) \times P\left(q_4 \mid q_3\right) \\
\times P\left(q_3 \mid q_4\right) \times \cdots \times P\left(q_5 \mid q_2\right)
\end{split}
\end{equation}

\noindent Finally, the $x_{th}$ order Markov model with the smallest cost is selected. We calculate the required state transition matrix from the Markov model with the order selected.

\subsubsection{Discover workload patterns}

After obtaining the state transition matrix and the Markov model order $x$, we proceed to determine whether an SQL sequence is a pattern by following WI~\cite{tran2015oracle}. Specifically, given a threshold $\theta$, $q_1 \ldots q_x q_y$ is identified as a pattern if the state transition probability $\mathrm{P}\left(q_y \mid q_1 \ldots q_x\right) \geq \theta$.


\begin{example}
   Continuing Example~\ref{example:oracle1}, we set the threshold $\theta$ to $0.7$ and the Markov model order to $1$. We first initialize the pattern as $q_1$. Since $P\left(q_2 \mid q_1\right)=\frac{1}{2}<\theta$, $q_1$ is returned as pattern. Next, we reset the new pattern as $q_2$ and continue the calculation until the new pattern is reset to $q_4$, where $P\left(q_3 \mid q_4\right)=\frac{3}{4}>\theta$. As a consequence, we update the pattern as $q_4 q_3$. Since $P\left(q_4 \mid q_3\right)=\frac{2}{3}<\theta$, $q_4 q_3$ is returned as a pattern. We repeat this procedure until reaching the end of the sequence.
\end{example}



\section{\Offlineori{}}
\label{sec:offline}

In this section, we detail the implementation of \Offline{}, which is employed to train parameters offline.

\subsection{\Label{}}
\label{sec:label}
As discussed in Section~\ref{sec:online}, it is necessary to classify the queries stored in SQL query log into different \emph{business groups}. To achieve this, we need to collect labels to train a classifier.
We divide label collection into two parts: label collection for public data, and label collection for industrial data. Note that, in order to avoid the heavy back prop operations caused by the update of the foundation model, we do not fine-tune FMs.

\subsubsection{Label collection for public data}
It is uncommon for public datasets to log different businesses into one query log. This is because the currently available datasets are obtained from relatively simple business logic, and do not meet the requirement of large-scale complex business. To this end, we propose a generated dataset by fusing multiple public datasets (cf. Section~\ref{sec:experiment}), 
whose ground truth label is obviously available.

\subsubsection{Label collection for industrial data} We aim to protect user privacy when collecting labeled SQL queries from industry applications. This implies that we should minimize the labels we collect without compromising classification accuracy.

We provide users with a unique ID for each business group. 
Requests with the unique ID may pass through another API that is encapsulated by the cloud database interface, which is generally non-transparent to users.
This implies that users only need to pass this unique ID as an extra data field when logging in to the cloud database. With the unique ID, we provide three options for users to label their business groups.

\noindent
\textit{\textbf{Random Sample.}} When the user queries the cloud database, the system will send the query with a probability $P_{L}$, where $P_{L}$ is a configurable parameter. 
The default value of $P_{L}$ is $0.01$, indicating that 1\% of the total requests are marked as training labels to classify the business groups. 

\noindent
\textit{\textbf{Manually Labeling.}}
Users can flexibly choose the queries to be labeled, by informing the system of their preferences. This preserves the user privacy. 

\noindent
\textit{\textbf{Hybrid.}}
When employing random sampling, users are allowed to manually specify the type of SQL query that is prohibited from accessing as the training label. Specifically, this is achieved by setting a flag when querying the SQL statement. The flag overwrites the global probability $P_{L}$ so that the queries with the flag will not be sampled by the system. This setting aims to balance the label quality and user privacy. 

When \emph{Random Sample} and \emph{Hybrid} are adopted, the classifier model is trained on a weekly basis by default and the frequency of updating the model is configurable. After training, the labeled SQL queries are discarded to prevent possible data/privacy leakage. When \emph{Manually Labeling} is adopted, the classifier model is trained only when the user requests the model to be updated.

\subsection{Classifier Model Training}

The parameters involved in 
\XGB{} should be trained accurately, which is a prerequisite for having \XGB{} perform classification efficiently.
To achieve this, we need to carefully select the training data. It can be collected from  \emph{business groups} who have executed massive SQL queries on the cloud database server. Note that, in order to protect user privacy, we aim to minimize the number of labels we collect. We adopt XGBClassifier~\cite{XGBoost16} as the classifier model. However, \Method{} also accommodates other classifier models (cf. Section~\ref{sec:experiment}).

As the size of the input data is extremely large, the model training follows a mini-batch strategy. Specifically, we separate the data by the `timestamp' feature and train the model in small batches. 


\section{\Optimizeori{}}
\label{sec:optimize}

\begin{figure}[t]
\includegraphics[width=.45\textwidth]{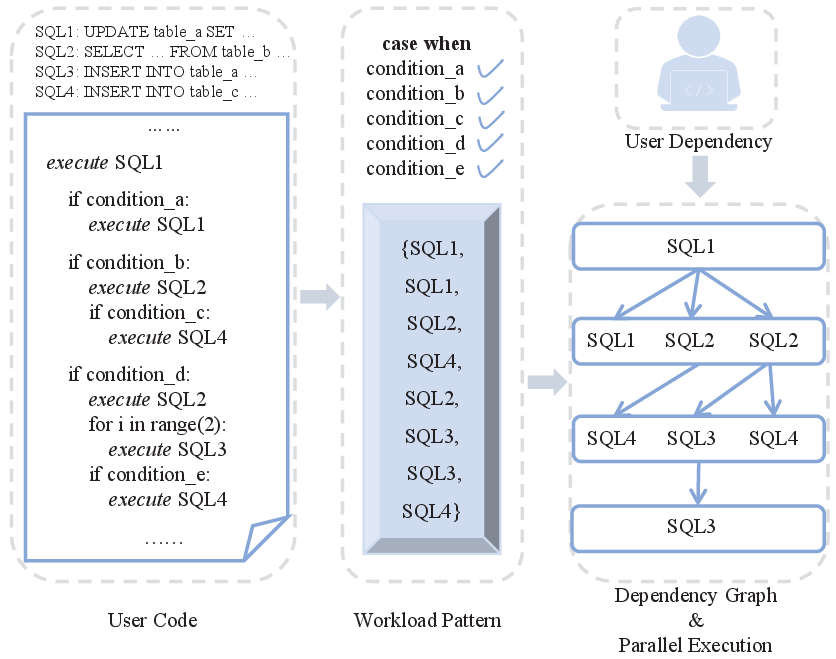}
\vspace{-4mm}
\caption{An example of optimizing query code.}
\label{fig:dep2}
\vspace{-6mm}
\end{figure}

As an application of \Method{}, \Optimize{} analyzes the workload patterns discovered by \Online{} and provides optimization strategies for cloud database users to improve their business logic codes.
Optimizing queries based on mined workload patterns can fundamentally optimize the business, which is more valuable than optimizing based on the entire workload extracted from logs directly.

In general, the workload patterns from one business group typically have two origins. The first is the pattern within one piece of code. For example, a function that first executes the "SELECT" command on the "user" table to select requested user IDs, and then iteratively "SELECT" and "UPDATE" the followers of the user. This creates a pattern of "SELECT uid" $\rightarrow$ "SELECT follower" $\rightarrow$ "UPDATE follower" $\rightarrow$... . The second is the pattern generated from the business group, which is the remote procedure call (RPC) from different micro-services. The reason is that the modern implementation of business logic follows the micro-service architecture~\cite{nadareishvili2016microservice}, which enables back-end programmers to reduce code coupling. This strategy is adopted by most large tech companies. This motivates us to take advantage of the analysis of this strategy. Specifically, in \Optimize{}, micro-services call each other to form a directed acyclic graph (DAG) with a call relationship in a user's request.

In both of the above situations, dependencies exist within the codes querying the cloud database, either intentional or not intentional. 
Furthermore, the dependencies are hidden inside the patterns discovered from the SQL query log corpus. 
For SQL queries without dependencies in-between each other, executing them in parallel is obviously more efficient. Since a pattern is a sequence of queries that occurs frequently, optimizing the execution of patterns is more significant than that of infrequently occurred queries.
Hence, our goal is to analyze the workload patterns and figure out strategies for users to execute SQL queries in parallel (as shown in Fig.~\ref{fig:dep2}). To achieve this, we represent the dependencies as a graph and propose \Topo{} for constructing the dependency graph within queries. Based on this, we provide optimization based on the constructed graph.

\lstnewenvironment{queryl}[1][] 
{\lstset{basicstyle=\fontsize{7}{6}\selectfont\ttfamily,frame=shadowbox,escapechar=`,linewidth=8cm, #1}}
{}

\begin{figure}[t]
\begin{queryl}[linewidth=0.45\textwidth]
// business-based dependency
if ("SELECT count(*) FROM users"==10){
    items = "SELECT ... FROM items"
    shops = "SELECT ... FROM shops"
    // block-based dependency
    "UPDATE users SET ...  WHERE ..."
    ...
}
\end{queryl}
\vspace{-4mm}
\caption{An example of query code.}
\label{fig:dep}
\vspace{-2mm}
\end{figure}

\subsection{\Topo{}}
\subsubsection{Dependency graph construction}
Two types of dependencies are related to constructing the dependency graph. The first type (namely \emph{block-based dependencies}) is the dependencies that are trivially inferred from the SQL query text. For example, if there is one "SELECT" operation after an "UPDATE" operation, and the two queries are applied to the same table, then the "SELECT" should not be executed until the "UPDATE" is completed. Intuitively, this type of dependency can be directly derived from the data. 
The second type of dependency (namely \emph{business-based dependencies}) is the business logic-related dependency and is not accessible for the cloud database. Example~\ref{example:unknown_dep} gives an example.

\begin{example}
    \label{example:unknown_dep}
    Assume that the user has written some codes starting with an "if" statement, which is to check whether a "SELECT" statement returns any possible value (as shown in Fig.~\ref{fig:dep}). If it does, the program will run two "SELECT"s to check further requirements, and then run an "UPDATE" to update the table.
\end{example}

Example~\ref{example:unknown_dep} suggests that the first "SELECT" blocks the followed-up operations. However, it cannot be detected by the cloud database, as it is not a blocking query. Since it is hard to detect this type of dependency by using only the limited knowledge stored in the cloud database, we provide the users an interface, by which they can input such dependency and we can then collect it.

Next, we introduce the generation of 
dependency graph from query patterns. Considering the block-based dependencies, it is essential to detect the \emph{blocking queries}. Such queries may block followed-up queries to execute, e.g., an "UPDATE" query is a blocking query. Different blocking queries have different scopes of blocking. For example, the DDL (Data Definition Language, e.g., "CREATE", "ALTER") queries generally have larger scopes than the DML (Data Manipulation Language, e.g., "INSERT", "UPDATE") queries. To construct the graph, we iterate the pattern. For each scope (database/table), we maintain a map of the current blocking query. If the query is within a scope (e.g., is querying a specified table), we add one dependency edge between the blocking query of this scope and the current query. If the query itself is a blocking query, we update the corresponding scope in the map. For those business-based dependencies, we provide interfaces for users to define their specified dependencies.

\subsubsection{Dependency-graph-based optimization}

After the dependency graph $G$ is built, \Topo{} performs a breadth-first search on $G$ to determine the optimal execution order of queries. Once the optimization result is generated, users receive a list of queries that can be executed in parallel in the same order. In the case of incorrect results, users can manually add dependencies and the system will update the order accordingly. By following the suggestions provided by the system, users can further optimize their code for improved performance.

\newcommand{\NCases}{\MRED{168}}
\newcommand{\NInstances}{\MRED{36}}
\newcommand{\NMasterInstances}{\MRED{31}}
\newcommand{\NSlaveInstances}{\MRED{5}}
\newcommand{\NQueries}{\MRED{9.4 billion}}
\newcommand{\NTemplates}{\MRED{77450}}
\newcommand{\TotalTimePeriod}{\MRED{1653 minutes}}
\newcommand{\MeanCPUCores}{\MRED{15.9}}
\newcommand{\MeanMemory}{\MRED{87.9GiB}}
\newcommand{\AvgTemplatePerCase}{\MRED{3357}}

\begin{table*}[t]\small
    \caption{Comparison results.}
    \label{tab:overall_result}
    \vspace*{-4mm}
    \resizebox{\linewidth}{!}{
    \begin{tabular}{c|c|ccccc|ccccc|ccccc}
\toprule
  &     & \multicolumn{5}{c|}{AQL-N}      & \multicolumn{5}{c|}{AQL-L}            & \multicolumn{5}{c}{OSQL}  \\ \cline{3-17} 
\multirow{-2}{*}{\textbf{Setting}} & \multirow{-2}{*}{\textbf{Method}} & {\begin{tabular}[c]{@{}c@{}}Pattern\\ Precision\end{tabular}} & \# & F1      & Latency & Time   & {\begin{tabular}[c]{@{}c@{}}Pattern\\ Precision\end{tabular}} & \# & F1      & Latency & Time     & {\begin{tabular}[c]{@{}c@{}}Pattern\\ Precision\end{tabular}} & \# & F1      & Latency & Time  \\ \hline
-    & WI    & 0\%   & 0  & -       & 0.00302 & -      & 0\%   & 0  & -       & 0.00335 & -        & 3.92\%      & 2  & -       & 0.35479 & -     \\ 
-     & WI-tid      & 15.24\%     & 25 & -       & 0.00254 & -      & 0.79\%           & 25 & -       & 0.00487 & -        & -     & -  & -       & -       & -     \\ 
-   & WI-kmeans   & 25.37\%   & 17 & -       & -  & 2007.80   & 17.39\%     & 20 & -       & -  & 9899.11    & 20.00\%   & 16  & -       & -  & 285.41  \\ \hline  E         &      & 86.54\%     & 45 & 99.21\% & 0.00246 & 169.45 & 83.78\%     & 62 & 99.13\% & 0.00253 & 769.23   & 81.72\%      & 76 & 91.71\% & 0.27956 & 7.50   \\  L        &       & 88.00\%     & 44 & 99.39\% & 0.00246 & 290.08 & 84.00\%    & 63 & 99.55\% & 0.00258 & 1,904.86 & 90.63\%    & 87 & 93.98\% & 0.26061 & 16.05 \\  N   & \multirow{-3}{*}{AWM}    & 89.58\% & 43 & 99.45\% & 0.00247 & 402.44 & 86.49\%    & 64 & 99.60\% & 0.00260  & 2,643.84 & 90.91\%    & 90 & 94.10\% & 0.25213 & 28.07 \\ 
\bottomrule
\end{tabular}
}
\begin{tablenotes}
\footnotesize
    \item[1]$^{*}$F1 is to measure the performance of \XGB{} and Precision is to measure the overall pattern discovery performance.
\end{tablenotes}
\vspace*{-2mm}
\end{table*}

\section{Experimental Evaluation}
\label{sec:experiment}

 We report on extensive experiments aimed at evaluating the performance of \Method{}.

\subsection{Experimental Setup}
\label{subsec:experiment_setup}
\textit{\textbf{Experimental dataset.}}
We use two real-life datasets (\alione{} and \alifour{}) and one synthetic dataset (\opendata{}).
\begin{itemize}[topsep=0pt,itemsep=0pt,parsep=0pt,partopsep=0pt,leftmargin=*]
    \item \textbf{\alione{}.} It is a normal query log of the Alibaba Cloud database.
    It contains 941K queries, of which the number of SQL templates is 184. The total query response time of \alione{} is 683.22 seconds.
    
    \item \textbf{\alifour{}.} It is a large query log of the Alibaba Cloud database.
    It contains 4.5M queries, of which the number of SQL templates is 205. The total query response time of \alifour{} is 2,512.59 seconds. 
    
    \item \textbf{\opendata{}}.
   \Method{} is designed to discover workload patterns from multiple sources. However, no public integrated dataset is available,  
      Thus, 
      we synthesize a dataset by fusing four public datasets: StackOverflow~\cite{iyer2016summarizing}, IIT Bombay~\cite{chandra2015data}, UB dataset~\cite{kul2018similarity}, and PocketData~\cite{kennedy2015pocket}.
    Specifically, we randomly sample queries from the four datasets and then mix them into \opendata{}.
     \opendata{} contains 4,403 queries, of which the number of SQL templates is 4,308.
    Since the four public datasets (StackOverflow, IIT Bombay, UB dataset, and PocketData) only provide the query statement texts but not the query execution feature, the total query response time of \opendata{} is unknown.

\end{itemize}

\alione{} and \alifour{} from Alibaba Cloud are \emph{diversified} and \emph{complicated}.
\emph{Diversified} means that they contain not only a large number of entries but also a variety of information such as SQL text and execution feature metrics. 
\emph{Complicated} means that they contain query entries from different business logic and database instances. We collect the SQL data from a certain application of the Alibaba Cloud database as the ground truth, in order to test the performance of \Method{} on \alione{} and \alifour{}. On the other hand, \opendata{} only contains a small number of queries with the repetition rates of templates being low.
This implies that most patterns of \opendata{} have shorter lengths. Since no ground truth is available, we collect data from each source of \opendata{} and apply \Oracle{} to generate ground truth patterns. 

\noindent
\textit{\textbf{Evaluation metrics.}}
We study the performance of \Method{} in terms of \emph{Effectiveness of pattern mining}, \emph{Effectiveness of \XGB{}}, and \emph{Efficiency of \Method{}}.
\emph{Effectiveness of pattern mining}  is measured by the number of correctly discovered patterns (\# of pattern) and precision (Precision).
\emph{Effectiveness of \XGB{}} is measured by F1-Score(\claacc{}).
    \emph{Efficiency of \Method{}}  is measured by logging the online serving latency of \Online{} (Latency) and the offline training time of \Offline{}  (Time).

\noindent
\textit{\textbf{Baselines.}}
We compare \Method{} with three baselines: WI~\cite{tran2015oracle}, WI-tid, and WI-kmeans.
WI mines the patterns of SQL context based on a Markov chain-based method, which is the state-of-the-art work for workload pattern discovery.
However, WI is designed for mining patters with a single business logic, and thus cannot be directly applied to discovery patterns in cloud databases (cf. Section~\ref{sec:intro}). 

To attain a fair comparison, we provide two variants of WI: WI-tid and WI-kmeans, both of which adapt WI to cloud databases.
WI-tid classifies the queries of \alione{} and \alifour{} using the \textit{tid} (i.e., the \underline{t}hread \underline{id} of the program to access the database) attribute of the data source. This way, queries from multiple sources can be distinguished according to the characteristic of the business logic. After classification, WI-tid exploits WI to discover patterns for each category respectively.
WI-kmeans performs classification in an unsupervised way, where the data is automatically categorized.
Specifically, WI-kmeans first adopts K-Means to cluster data, and then discovers patterns for each cluster with WI.

\noindent
\textit{\textbf{Implementation details.}}
 We set the maximum order of the Markov model to 1 and $\theta$ to 0.77 for all baselines.
  When implementing WI-kmeans, we set the number of clusters to 5 for \alione{} and \alifour{}, and 15 for \opendata{}.
  We use the encoded query features obtained from \XGBInput{} to cluster the queries for \alione{} and \alifour{}. Since the query execution feature is unavailable, 
  we apply \BERT{} to the query texts and take the embeddings as the clustering input. 
  All the baselines do not have a classification model. 
    Thus, we do not report \claacc{} and \clatime{} for them. Since WI-kmeans runs clustering on the whole dataset, it is infeasible to apply it to online scenarios. 
    Hence, we do not report the \latency{} for WI-kmeans. Moreover, \clatime{} of WI-kmeans refers to the end-to-end workload discovery time.
We set three training set ratios
(proportion of the data used for training) for each dataset, respectively, when studying the performance of \Method{}. 
The lowest training set ratio is denoted as E; the highest training set ratio is denoted as N; and the middle training set ratio is denoted as L.
 For \alione{} and \alifour{}, we choose 1\% (E), 5\% (L), and 10\% (N) of the total data of  \alione{}, \alifour{}, and \opendata{} as training sets, respectively.  Here, the highest training ratio (10\%) is below the commonly used settings of existing classification tasks~\cite{XGBoost16}; while the lowest training ratio (1\%) is extremely smaller than the one adopted by most of the classification tasks~\cite{XGBoost16}.
  For \opendata{},  we choose 20\% (E), 40\% (L), 60\% (N) of the total data of  \alione{}, \alifour{}, and \opendata{} as training sets, respectively. Here, we use a larger proportion of training data because the query execution features of \opendata{} are unavailable.
    We employ the max pooling method and a batch size of 512 for the \BERT{}'s foundation model. We select multi:softmax as the loss function and mlogloss as the evaluation function for \XGB{}. We use the same setting as WI when implementing \Oracle{}.
    We set the number of batches to 10 for \XGBInput{} of \alifour{}.

\vspace{-2mm}
\subsection{Comparison Study}
\label{sec:main_results}
We compare \Method{} with WI~\cite{tran2015oracle}, WI-tid and WI-kmeans
in terms of effectiveness and efficiency. 
 \Method{}, WI and WI-kmeans are performed on \alione{}, \alifour{}, and \opendata{}, while WI-tid is performed only on \alione{} and \alifour{}. This is because WI-tid conducts classification by exploiting the \textit{tid} attribute (cf. Section ~\ref{subsec:experiment_setup}), which is missed in \opendata{}.
 Table~\ref{tab:overall_result} shows the comparison results.

\begin{figure*}[t]
\centering
    \vspace{-6mm}

\includegraphics[width=0.65\textwidth]{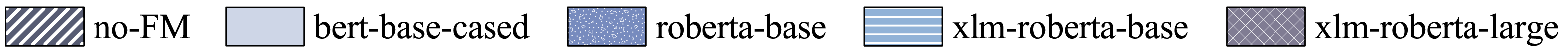}\\
\vspace*{-5mm}
\hspace*{-3mm}

\subfigure[\pacc{}]{
 \includegraphics[width=1.48in]{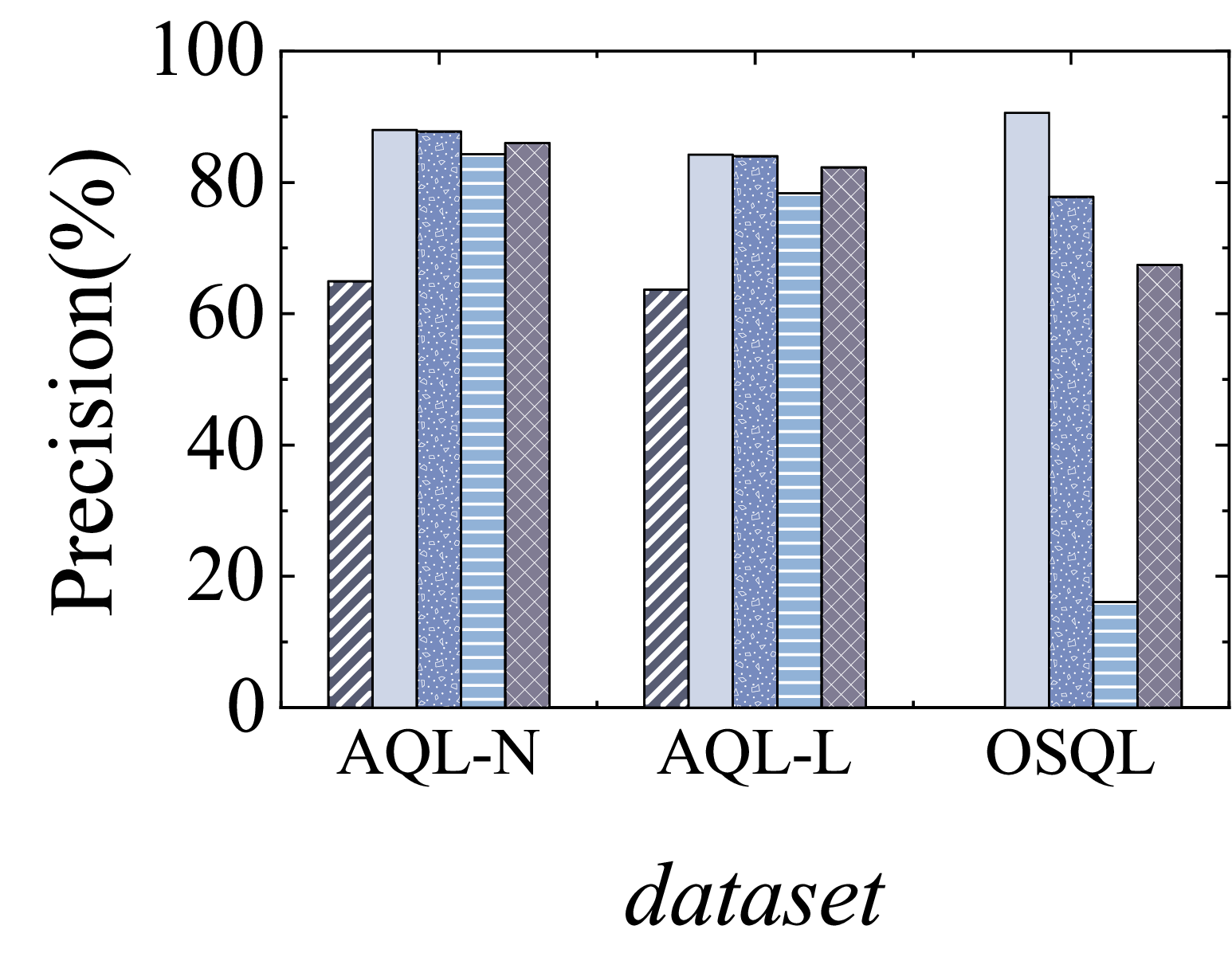}
}\hspace*{3mm}
\subfigure[\pnum{}]{
 \includegraphics[width=1.48in]{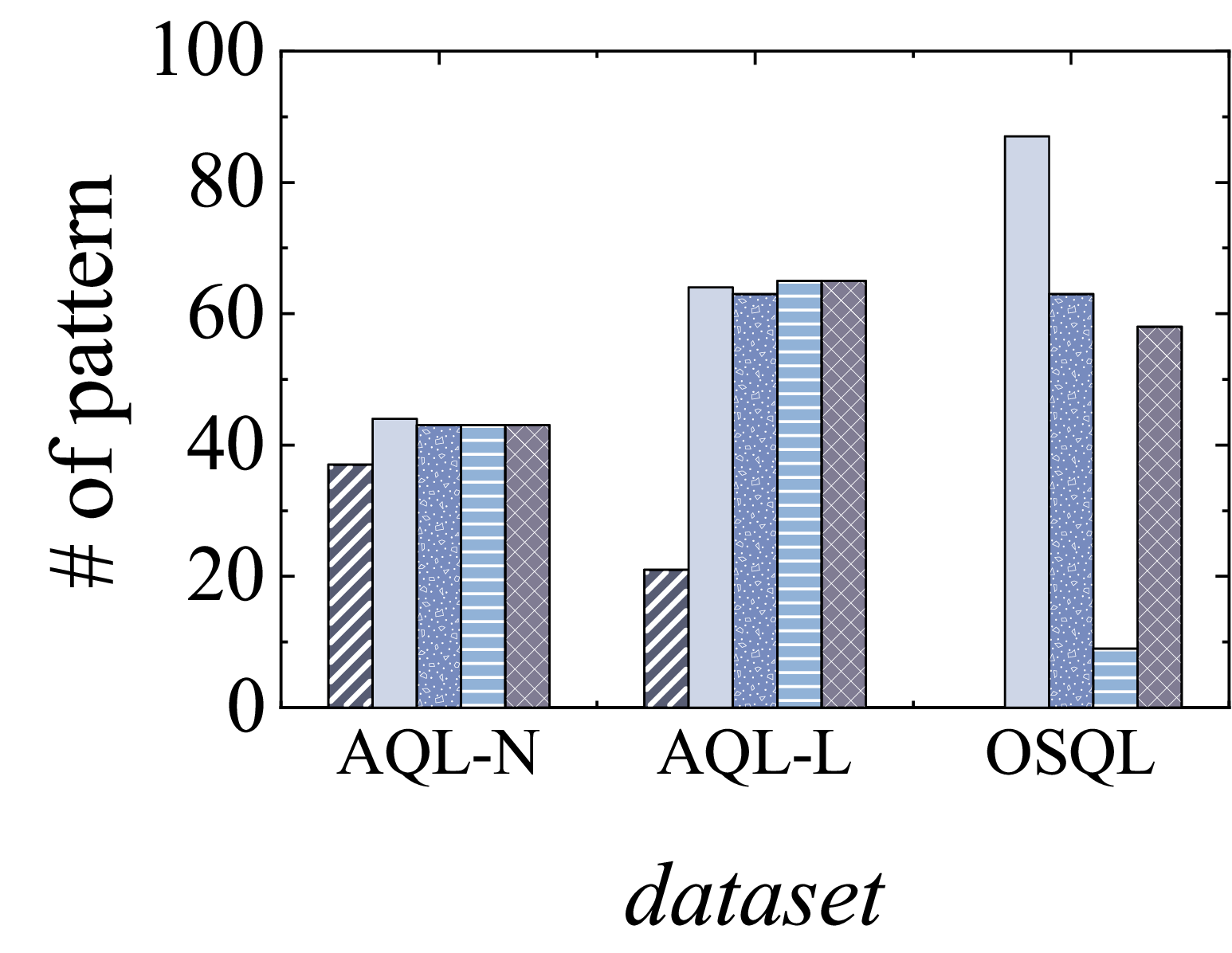}
}\hspace*{3mm}
\subfigure[\claacc{}]{
 \includegraphics[width=1.48in]{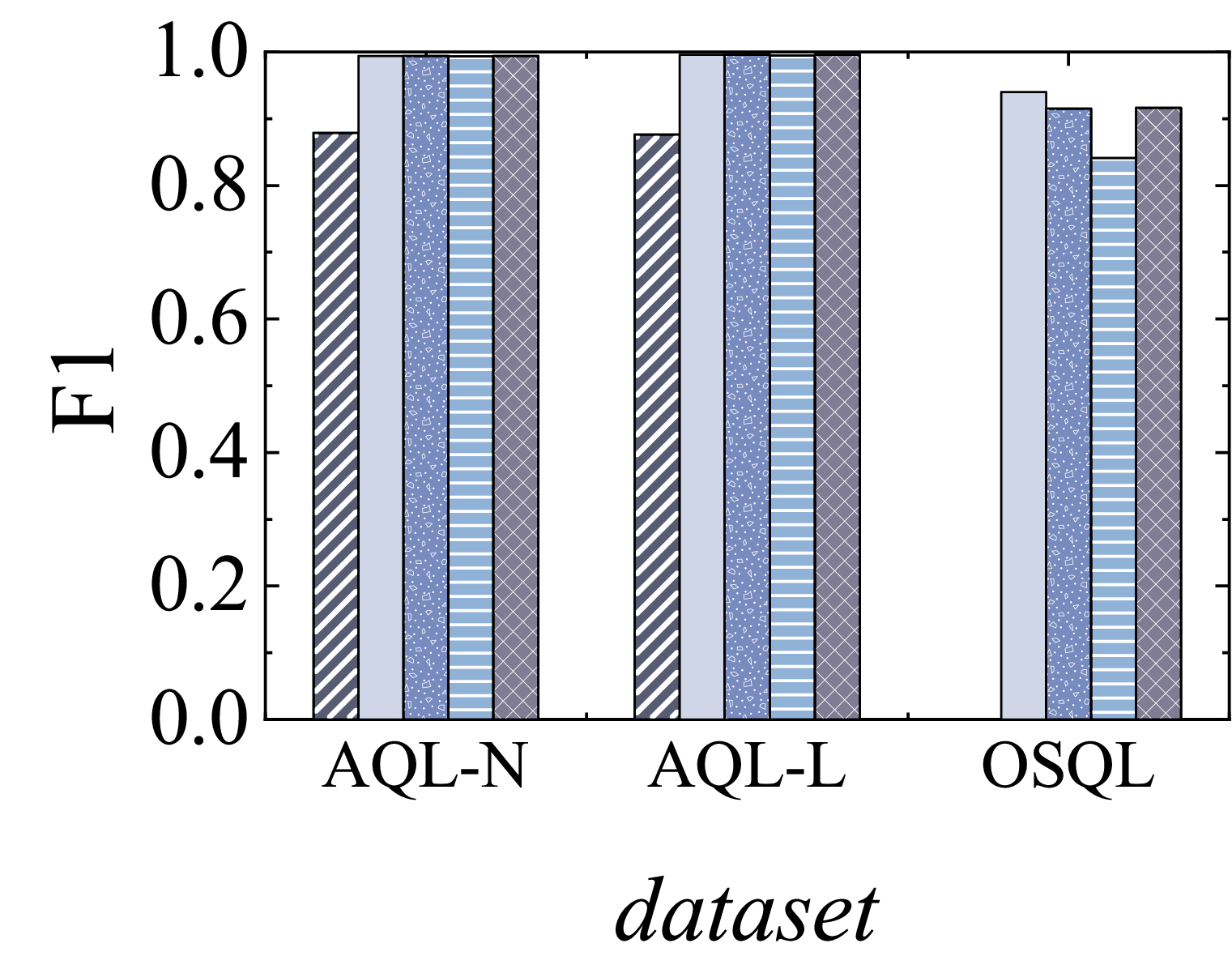}
}\hspace*{3mm}
\subfigure[\latency{}]{
 \includegraphics[width=1.48in]{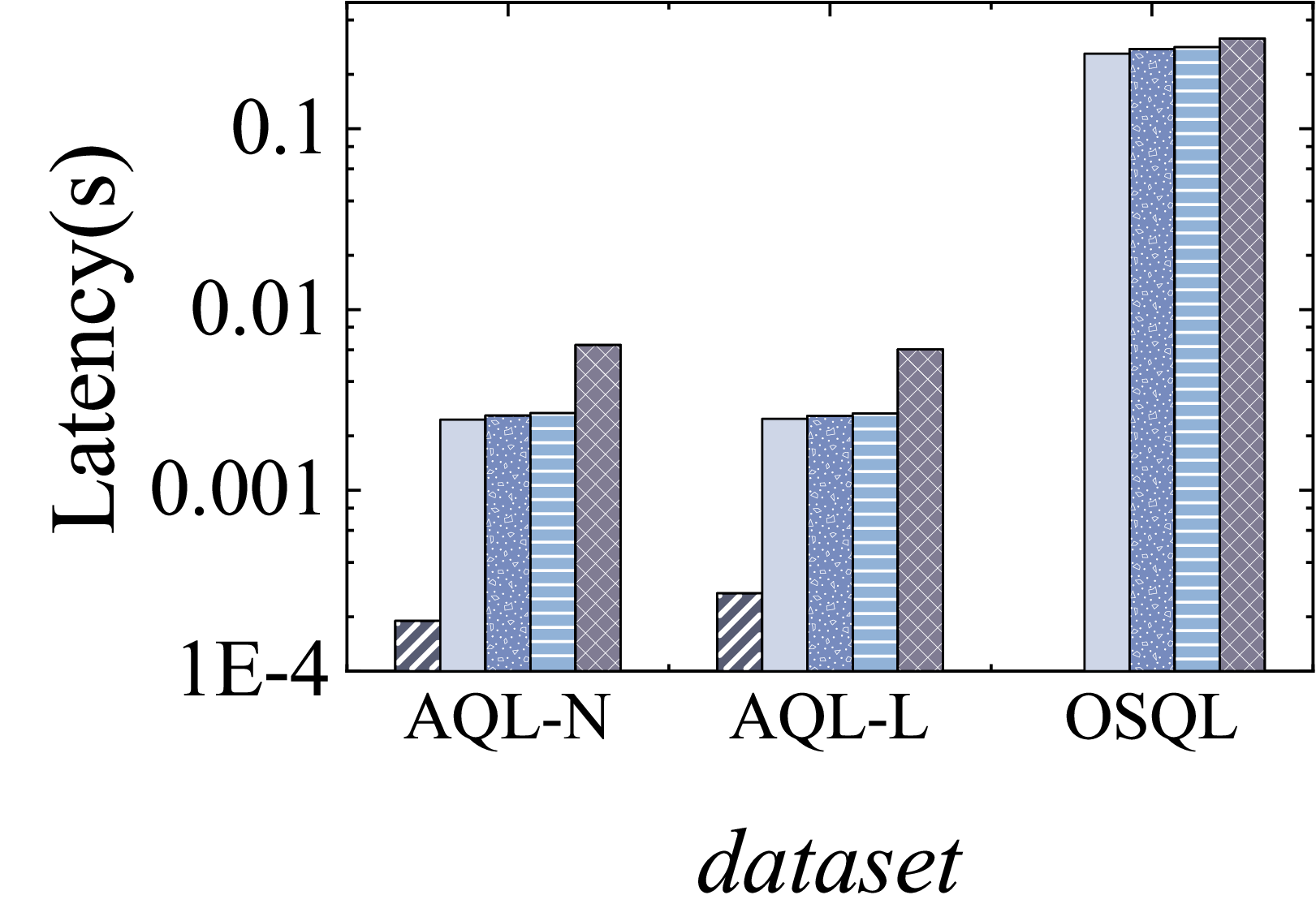}
}
\vspace*{-5mm}
\caption{The results of varying foundation models.}
\vspace*{-3mm}
\label{fig:lm_analysis}
\end{figure*}

\begin{figure*}[t]
\centering
\vspace*{-1mm}
\includegraphics[width=0.51\textwidth]{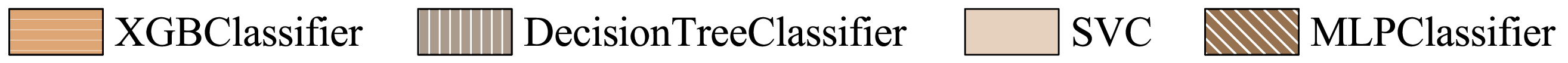}\\
\vspace*{-5mm}
\hspace*{-3mm}

\subfigure[\pacc{}]{
 \includegraphics[width=1.48in]{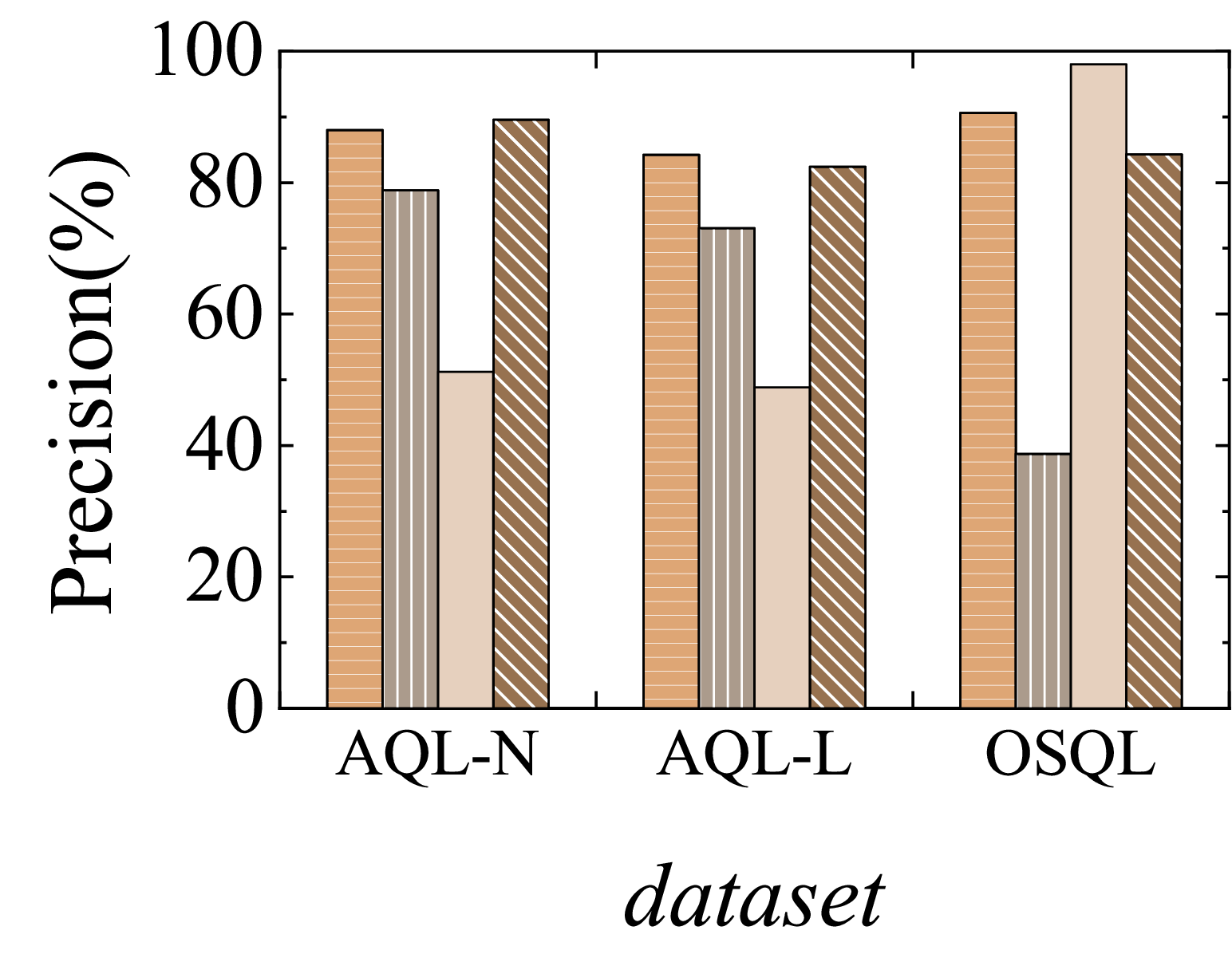}
}\hspace*{3mm}
\subfigure[\pnum{}]{
 \includegraphics[width=1.48in]{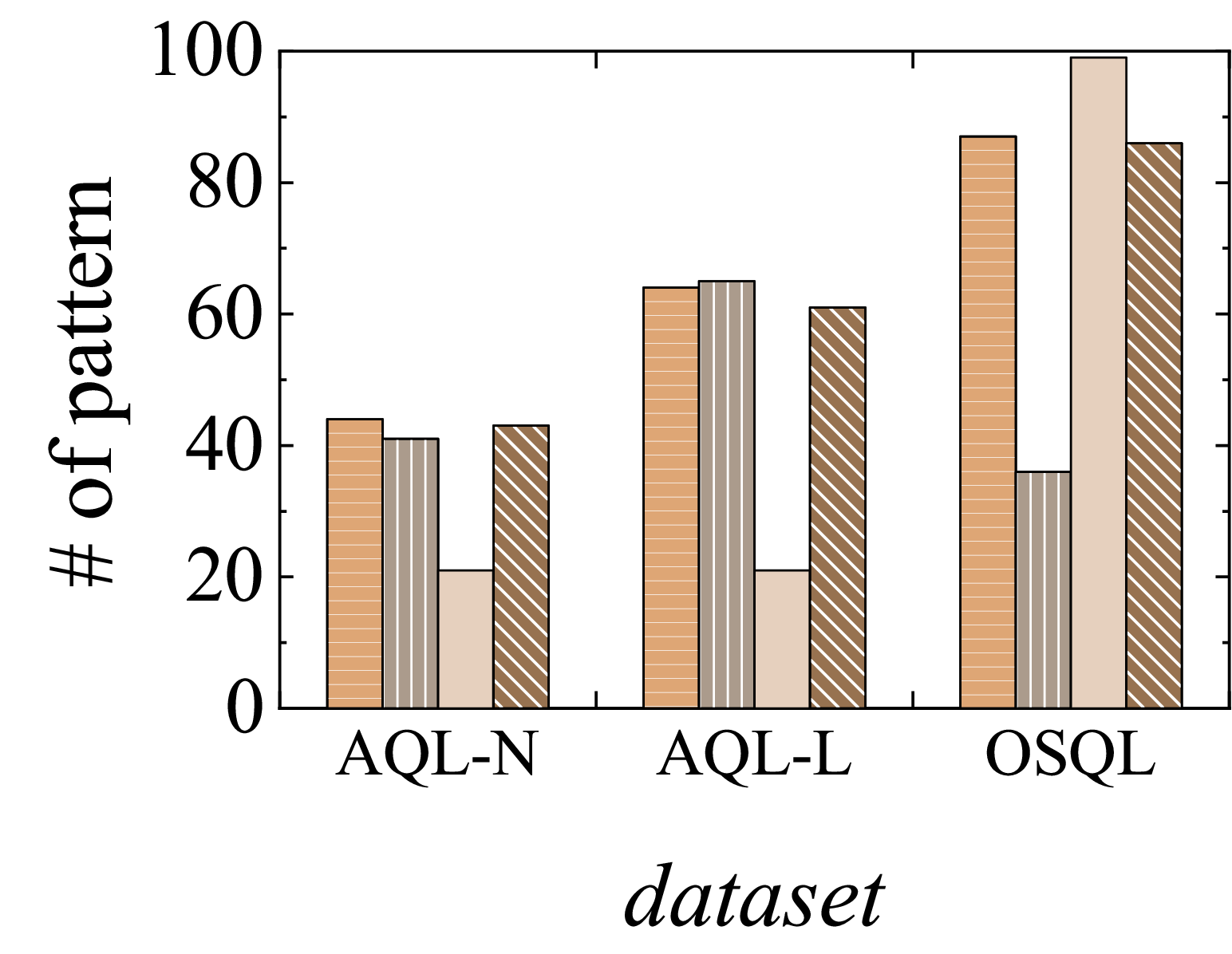}
}\hspace*{3mm}
\subfigure[\claacc{}]{
 \includegraphics[width=1.48in]{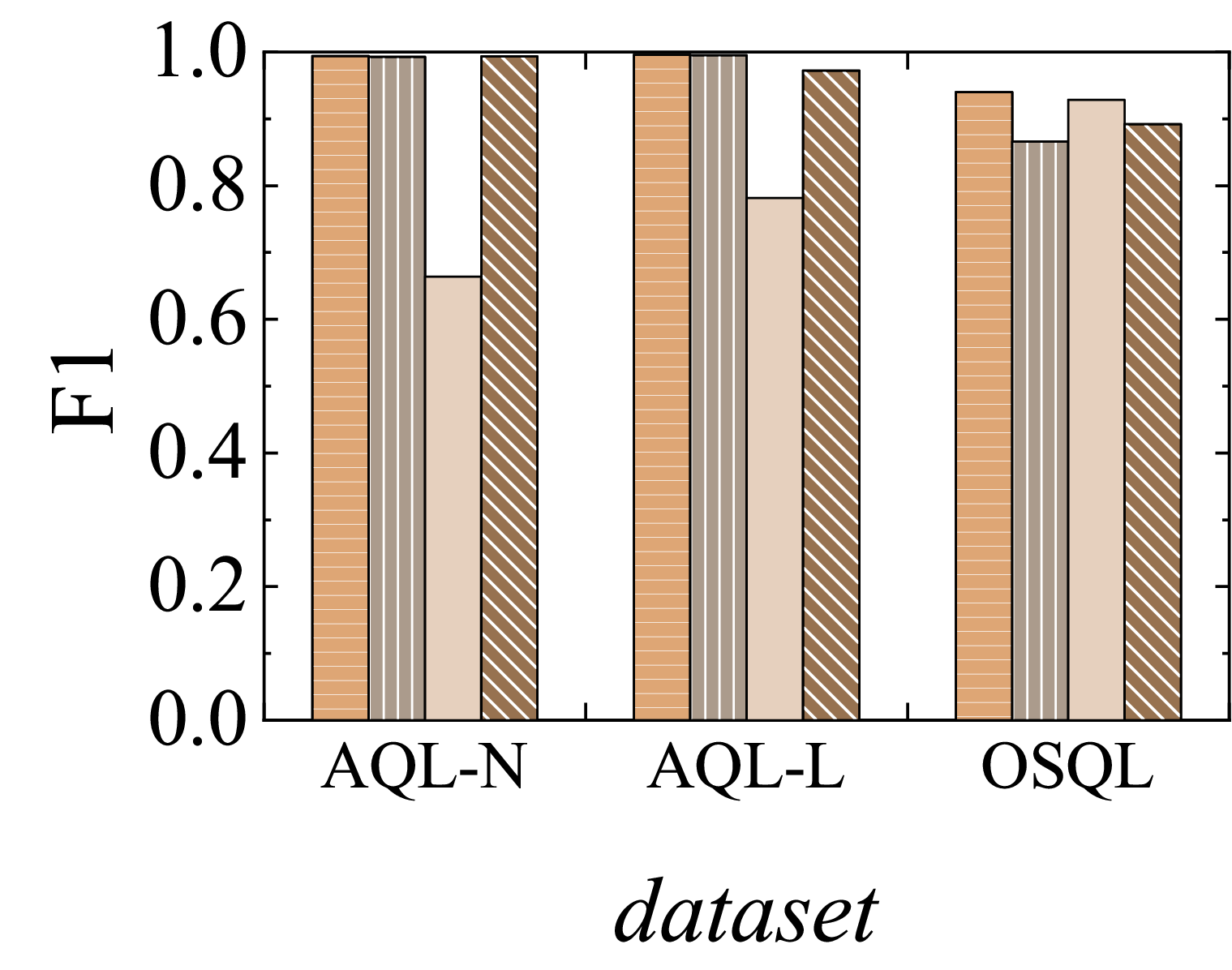}
}\hspace*{3mm}
\subfigure[\clatime{}]{
 \includegraphics[width=1.48in]{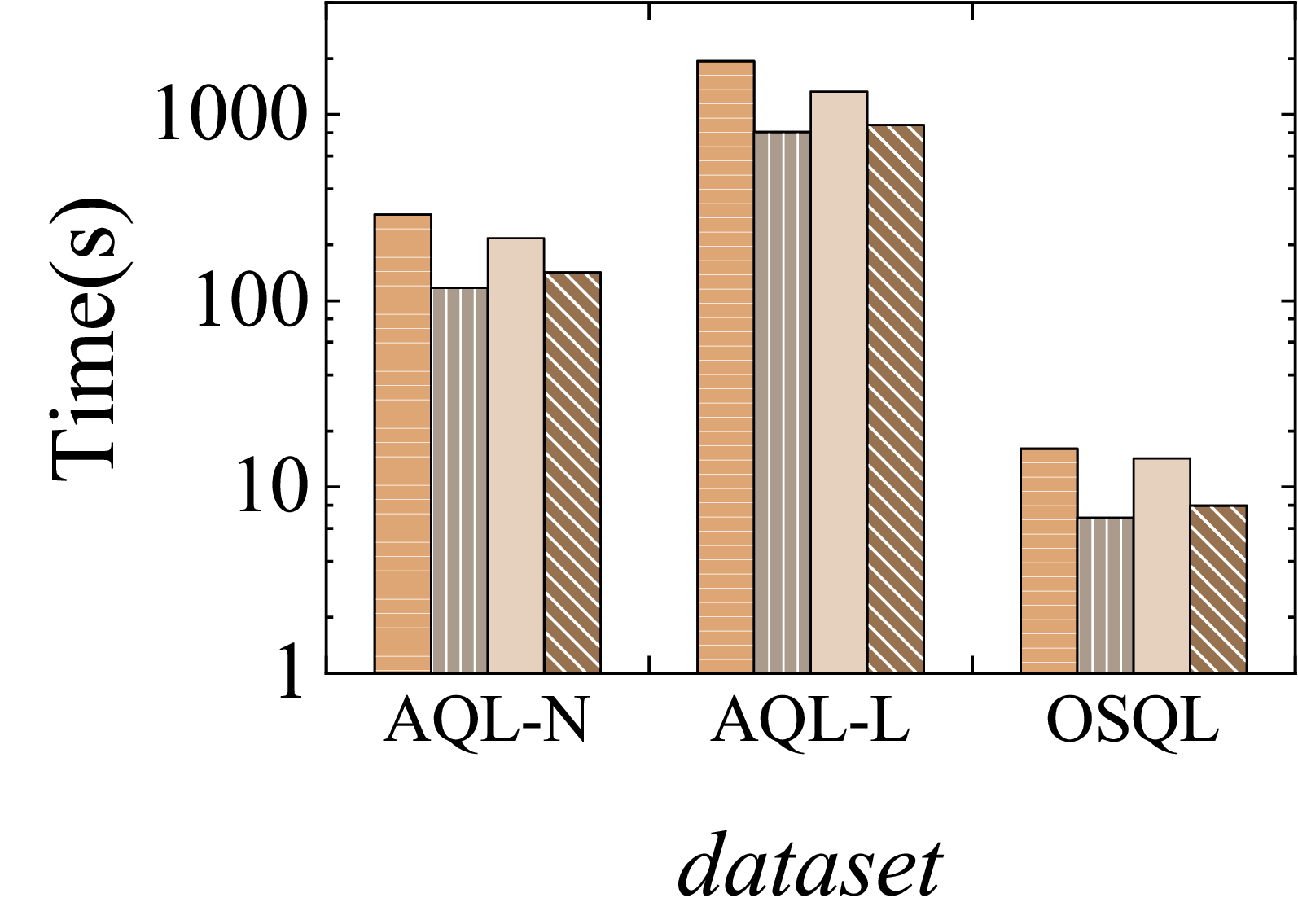}
}
\vspace*{-5mm}
\caption{The results of varying classifiers.}
\vspace*{-4mm}
\label{fig:classifier_analysis}
\end{figure*}

\subsubsection{Effectiveness Study.}
As observed in Table~\ref{tab:overall_result}, \Method{} outperforms three baselines significantly in terms of \pacc{} and \pnum{} on all datasets. First, Pattern Precision of \Method{} is more than 81\% and F1 is more than 91\% on all datasets. Specifically, \Method{} improves \pacc{} by more than 60\% compared to WI-kmeans, which achieves the highest \pacc{} among all baselines.
Next, patterns identified  by \Method{} (\# of the pattern) are 18 and 37 more than WI-tid on \alione{} and \alifour{}, respectively, and are 60 more than WI-kmeans on \opendata{}. These results highlight the effectiveness of \Method{}. This is attributed to the use of XGBClassifier~\cite{XGBoost16, ma2020diagnosing, yoon2016dbsherlock} model in \Offline{}, which takes into account data characteristics to more accurately mine patterns. Note that, both Pattern Precision and \# of the pattern of WI on \alione{} and \alifour{} is 0.
This is because WI is designed for a single database and small-scale data only (cf. Section ~\ref{subsec:experiment_setup}) and struggles when processing large-scale industrial data.

Third, the increase of training data leads to growths of \claacc{}, \pacc{} and \pnum{}, because more training labels lead to higher accuracy of training.
 However, even with the minimum training set ratios, \Method{} is able to outperform all baselines significantly. The reason is that \Label{} enables \Method{} to work with limited resources, where the user privacy is protected
  by  \Label{} (cf. Section~\ref{sec:label}).

\vspace{-2mm}
\subsubsection{Efficiency Study}
As shown in Table~\ref{tab:overall_result}, \Method{}'s \latency{} is significantly lower than the three baselines' on three datasets. Specifically, \Method{} reduces \latency{} by 2.7\% compared with WI-tid on \alione{} and reduces \latency{} by 22\% and 21\% compared with WI on \alifour{} and \opendata{}, respectively. Here, WI-tid achieves the lowest Latency among all baselines on \alione{} and WI achieves the lowest Latency among all baselines on \alifour{} and \opendata{}, respectively.
The above mentioned experimental results are mainly because (i) when processing large-scale data, all baselines generate large Markov models, resulting in a large time overhead; (ii) WI-kmeans performs clustering offline; and (iii) \Method{} reduces the sizes of Markov models by discovering patterns for each category respectively and separating offline training from online classification and workload pattern discovery. 

Next, \latency{} of all methods on \opendata{} is significantly higher than those of the other two datasets. This is because \latency{} largely depends on the number of SQL templates. As mentioned in Section ~\ref{subsec:experiment_setup}, \opendata{} has more SQL templates and a lower repetition rate of SQL queries than \alifour{} and \alione{}. This leads to higher time and space costs for selecting an appropriate Markov model in \Oracle{}.
Meanwhile, the latency of WI-tid on \alifour{} is higher than that on \alione{}. This reason is that when processing massive data, classifying by \textit{tid} tends to generate more repeated patterns, which increases the time cost.

Although it is true that \Method{} requires additional time for training the classification model offline, its time cost is still acceptable for processing millions of data. As the size of data used for training the classification model grows, the \clatime{} of \Method{} increases. 
However, \Method{} requires only 100 seconds to train the model, which is negligible compared to the total query time (about 1 hour). Note that, \clatime{} of WI-kmeans is much larger than \Method{} because it mines all workload patterns in an offline fashion.
 
\vspace{-2mm}
\subsection{Ablation Study}
\subsubsection{Analysis of foundation models}
\label{sec:lm_analysis} Fig. \ref{fig:lm_analysis} reports the effects of four foundation models, bert-base-cased~\cite{bert}, roberta-base~\cite{liu2019roberta}, xlm-roberta-base~\cite{conneau2019unsupervised} and xlm-roberta-large~\cite{conneau2019unsupervised} by applying them to \Method{} respectively. We set the training ratio of each dataset to the corresponding lowest one, i.e., 1\% on \alifour{} and \alione{}, and 20\% on \opendata{} (cf. Section~\ref{subsec:experiment_setup}). no-FM denotes the case where we do not use any foundation model. Note that, no-FM has not been studied on \opendata{} because SQL text is the only available information for the public datasets~\cite{iyer2016summarizing, chandra2015data, kul2018similarity, kennedy2015pocket} (cf. Section~\ref{subsec:experiment_setup}).

The results  suggest that using foundation models in \Method{} can improve its performance in terms of \claacc{}, \pacc{} and \pnum{}. Although the use of foundation models does incur some time cost, the improvement in performance is significant and acceptable (cf. Section~\ref{subsubsec:embedding_store} and Section~\ref{exp:time}). Next, the bert-base-cased model achieves the best performance on each dataset, indicating that it is more universal than other models. 
The xlm-roberta-base model performs the worst in \opendata{} while performing well on the other two datasets. 
This is because the training data of xlm-roberta-base model differs largely from \opendata{}. Meanwhile, we observe that when processing large-scale data, a small deviation in-between \claacc{} may indicate a huge gap in between the number of correctly discovered patterns.  Finally, xlm-roberta-large model's \latency{} is 2 times larger than other foundation models'.
The reason is that the size of xlm-roberta-large model is the largest among the three models, and thus results in the largest embedding dimension and longest time for preprocessing in \DataPreprocess{}.

\subsubsection{Analysis of classifier.}
\label{sec:classifier_analysis}
Fig. \ref{fig:classifier_analysis} shows the effect of using different classifiers. 
 As observed, XGBClassifier and MLPClassifier are more suitable for \alione{} and \alifour{}, while SVC is more suitable for \opendata{}. This is because \opendata{} have more SQL templates. On the other hand, \alione{} and \alifour{} only have around 200 unique templates and contain additional execution features which do not fit the SVC model. 
Moreover, according to the experimental results on \alifour{}, Time grows with the data size, because we employ batch processing to trade time for space.

\vspace{-2mm}
\subsection{Scalability Study}
\label{exp:time}

We study the scalability of the Embedding store and \Offline{} of \Method{} on \alione{} and \alifour{}. We remove the Embedding store (by applying FMs online to obtain the embedding) and \Offline{} from \Method{}, respectively. We denote the system without the Embedding store as \Method{}-Embedding store and the system without \Offline{} as \Method{}-\Offline{}. Then we compare them with \Method{}, respectively.
Note that, since directly calculating the embeddings of the whole dataset is infeasible, we randomly sample 2\% data of the total data and record \latency{} of \Method{}, \Method{}-Embedding store, and \Method{}-\Offline{} on it.

\begin{figure}[t]
\includegraphics[width=0.37\textwidth]{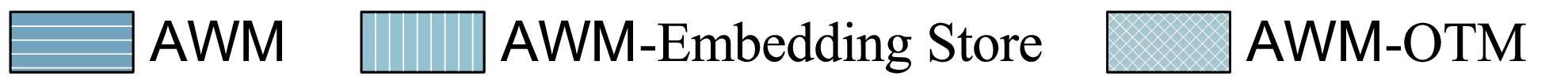}\\
\vspace*{-5mm}
\hspace*{-3mm}

\subfigure[Embedding Store Comparison]{
 \includegraphics[height=1.06in]{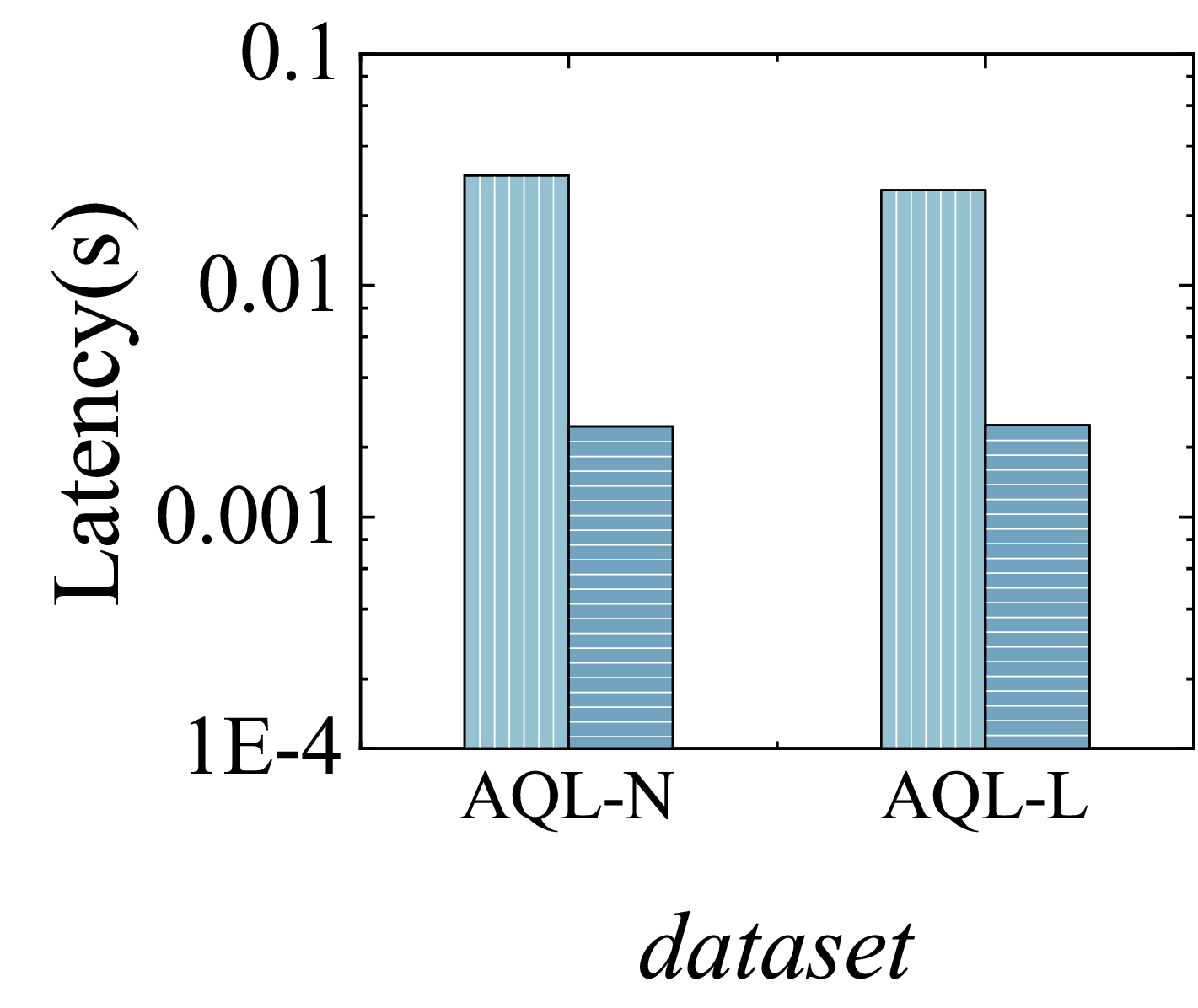}
}
\hspace*{4mm}
\subfigure[Offline Training Comparison]{
 \includegraphics[height=1.06in]{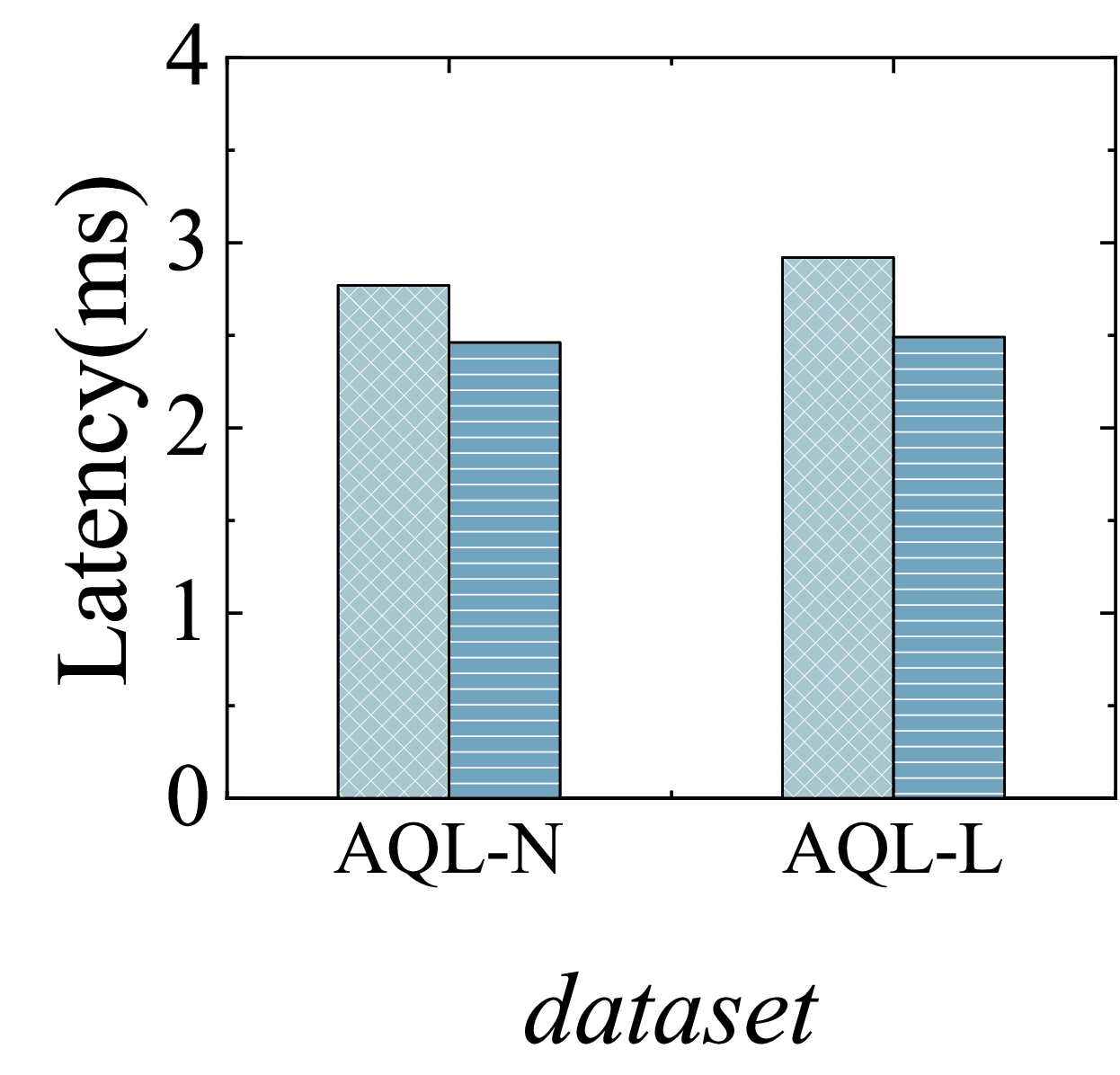}
}

\vspace*{-2mm}
\vspace*{-2mm}
\caption{The results of scalability study.}
\label{fig:ablation_analysis}
\end{figure}

The experimental results are reported in Fig.~\ref{fig:ablation_analysis}. First, as shown in Fig.~\ref{fig:ablation_analysis}a, \latency{} of \Method{}-Embedding store 
 is 10 times higher than that of \Method{}. 
This verifies that employing the Embedding store can greatly reduce the time for computing the embedding of massive data, thereby improving the efficiency of \Method{}. Next, as shown in Fig.~\ref{fig:ablation_analysis}b, \latency{} of \Method{}-\Offline{} is 1.15 times higher than that of \Method{}. This is because \Method{} trains data offline, which is more efficient.

\subsection{Study of \Optimizeori{}}
\label{sec:optimization_analysis}

We evaluate the effectiveness of \Optimize{}. 
We collect ten user cases in an internal software evaluation session. Each user case is a piece of code that contains queries. We divide the ten cases into two categories: (i) cases with loops and (ii) cases without loops.
We first apply \Method{} to the ten cases to identify the workload patterns and then use the patterns to generate optimization strategies with \Optimize{}. Next, we improve the codes with optimization strategies. 
We run each piece of original code and improved code 1,000 times, respectively, and record two metrics: the average execution time (denoted as mean) and the variance of execution time (denoted as std). We denote the situation where we run the original codes as pre-opt and the situation where we run the improved codes as post-opt.
 
Table~\ref{tab:optimizing_result} shows the query performance conparison of post-opt and pre-opt, respectively.  
 It is evident that \Optimize{} has a considerable impact on the query execution time. Specifically,
 it reduces the average execution time by 2.7 times. Moreover, it effectively decreases the variance of execution time. This indicates that employing an optimized parallel strategy improves the stability and predictability of query execution. Next, \Optimize{} is more beneficial for cases that contain loops.
The reason behind this observation is that patterns with more loops can execute a greater number of queries in parallel, which leads to better efficiency. The module can recognize such patterns and generate optimization strategies accordingly to leverage the parallelism and improve the performance further.

\begin{table}[t]\small
  \vspace{-3mm}
 \caption{Results of optimizing.}
 \label{tab:optimizing_result}
 \vspace*{-4mm}
 \resizebox{\linewidth}{!}{
 \begin{tabular}{c|cc|cc|c|c}
\toprule
\multirow{2}{*}{} & \multicolumn{2}{c|}{pre-opt} & \multicolumn{2}{c|}{post-opt} & \multirow{2}{*}{\begin{tabular}[c]{@{}c@{}}$\downarrow$mean\\ \end{tabular}} & \multirow{2}{*}{\begin{tabular}[c]{@{}c@{}}$\downarrow$ std\\\end{tabular}} \\ \cline{2-5}
 & mean   & std & mean   & std  & & \\ \hline
All cases   & 1.692  & 0.120    & 0.620  & 0.004 & 2.7 & 33.4   \\ \hline
No loop & 0.654  & 0.012    & 0.519  & 0.003 & 1.3 & 4.6    \\
With loop & 2.136  & 0.166    & 0.663  & 0.004 & 3.2 & 41.5   \\
\bottomrule   
\end{tabular}
 }
 \up\up
\end{table}

\section{Related Work}
\label{sec:related work}
We provide an overview of the related works, including automatic workload analysis, query optimization, automatic database analysis, and pre-trained foundation models.

\noindent \textit{\textbf{Workload analysis.}}
Analyzing and understanding workloads is essential for properly designing, provisioning and optimizing services~\cite{query2vec,doc2vec,PreQR,bert,tran2015oracle,deep2020comprehensive, oneprovenance2022, zhu2021kea}.
There is recent work on exploiting workload characteristics in cloud DBMSs from the query level~\cite{oneprovenance2022} and the system level~\cite{zhu2021kea}.
Although some work on workload characterization analyzes SQL logs~\cite{paul2021database,query2vec,PreQR,facilitating2020}, most of them focus on the interior of a single SQL statement.
Query2Vec~\cite{query2vec} implements a vector representation of SQL queries with natural language processing (NLP) methods,
 to support workload analysis tasks using a corpus.
PreQR~\cite{PreQR} improves BERT~\cite{bert} by taking into account the specific database information to learn query features and transforms human written texts into SQL queries.

As a branch of workload analysis, workload pattern mining aims to discover frequent patterns in query logs.
Oracle Workload Intelligence (WI)~\cite{tran2015oracle} proposes a Markov chain-based method for mining SQL context. 
However, WI cannot be applied to industry applications. This is because in such applications, queries from multiple business logics are loaded together into a query log store, making it difficult to distinguish them using WI. On the contrary, \Method{} 
employs a classifier to solve this problem. Although there are pattern mining studies targeting streaming data~\cite{giannella2003mining, jin2007frequent}, they are not applicable for discovering patterns from query logs.


\noindent \textit{\textbf{Query optimization.}}
Query optimization technologies are generally designed for optimizing single queries, by estimating the cost of query execution~\cite{SIGMOD19Index} or optimizing the query execution plan with deep neural networks~\cite{neo,bao} or  Monte Carlo Tree~\cite{learnedrewrite}. However, they cannot optimize multiple queries. 


Some studies investigate multi-query optimization from the perspective of batch processing~\cite{shared2014,sharedb2013,psidb2019,guided2019}, but none of them analyze the contextual characteristics of workload.
To tackle this issue, \Method{} exploits the discovered patterns to optimize query execution.

\noindent \textit{\textbf{Automatic database analysis.}}
Automatic analysis and optimization of databases include various functions such as tuning~\cite{VLDB09iTuned}, optimizing~\cite{VLDB11Profiling, bao, neo, learnedqueryoptim, learnedrewrite}, 
and workload management~\cite{VLDB16WiseDB}, which aim to improve the performance of database systems by addressing stability, scalability and etc. Some self-driving DBMSs~\cite{CIDR17SelfDriving, li2021opengauss} integrate these functions to automate the overall performance of database systems. However, they do not necessarily account for user behavior and business logic. \Method{}, on the other hand, focuses on analyzing and optimizing the database based on user behaviors and business logic through workload pattern analysis.

\noindent \textbf{\textit{Pre-trained foundation models.}}
Pre-trained foundation models (PFMs) have shown great success in various NLP tasks, by exploiting large amounts of unlabeled text data to learn common language representations ~\cite{bert, liu2019roberta, transformers}.  PFMs have been used in tasks such as sentence embedding~\cite{sbert}, matching~\cite{ge2021collaborem}, and knowledge graphs~\cite{EASY21, LargeEA22, ClusterEA, DualMatch, wu2023sea}. However, these models are not directly applicable in the database domain. The reason is that they are trained on web corpus that are significantly different from SQL queries. Although PreQR~\cite{PreQR} applies BERT to SQL statements for  tasks, such as cardinality estimation and text-to-SQL transformation, it is not designed to discover workload patterns from query logs. On the other hand, \Method{} leverages the benefits of PFMs and is tailored for the unique characteristics of SQL queries and database workloads.


\section{Conclusions}
\label{sec:conclusions}

In this paper, we propose a workload pattern discovery system \Method{} for large-scale industry-level workload analysis in real-time. \Method{} is a part of the Alibaba Cloud Database Autonomous Service (DAS).
Our system \Method{} (i) only collects data that is specified by users to protect user privacy; (ii) trains a classifier with few labels while achieve high accuracy; (iii) discovers the workload patterns logged from the user requests; and (iv) analyzes and provides strategies for optimizing those patterns.  
\Method{} is mainly composed of four modules: \DataPreprocess{}, \Online{}, \Offline{}, and \Optimize{}.
First, \DataPreprocess{} is applied to collect the streaming raw query logs, and encodes the raw query logs into high-dimensional feature embeddings with rich semantic contexts and execution features.
Next, \Online{} firstly classifies the encoded query logs by business groups, and then discovers the workload patterns effectively for each business group.
Meanwhile, \Offline{} collects labels that are shared by users and trains the classification model accordingly. 
Finally, \Method{} automatically provides clear code optimization strategies for cloud database users.
Extensive experiments have been conducted on one synthetic dataset and two real-life datasets (from Alibaba Cloud Databases). Specifically, the experimental results offer the insight that comparing with the state-of-the-arts, the accuracy of pattern discovery has been improved by 66\%; the latency of  online inferring has been improved by 22\%; and the optimization strategies offered by the system have improved the efficiency of query processing by 2.7 times. 



In the future, it is of great interest to discover the relation between database performance indicators and workload patterns. Mining workload patterns has important applications, such as query optimization. This paper focuses on a specific aspect of query optimization. However, workload patterns also hold promise for integration into online anomaly detection and root cause analysis.

\bibliographystyle{ACM-Reference-Format}
\bibliography{sample}


\begin{thebibliography}{59}


\ifx \showCODEN    \undefined \def \showCODEN     #1{\unskip}     \fi
\ifx \showDOI      \undefined \def \showDOI       #1{#1}\fi
\ifx \showISBNx    \undefined \def \showISBNx     #1{\unskip}     \fi
\ifx \showISBNxiii \undefined \def \showISBNxiii  #1{\unskip}     \fi
\ifx \showISSN     \undefined \def \showISSN      #1{\unskip}     \fi
\ifx \showLCCN     \undefined \def \showLCCN      #1{\unskip}     \fi
\ifx \shownote     \undefined \def \shownote      #1{#1}          \fi
\ifx \showarticletitle \undefined \def \showarticletitle #1{#1}   \fi
\ifx \showURL      \undefined \def \showURL       {\relax}        \fi
\providecommand\bibfield[2]{#2}
\providecommand\bibinfo[2]{#2}
\providecommand\natexlab[1]{#1}
\providecommand\showeprint[2][]{arXiv:#2}

\bibitem[\protect\citeauthoryear{{Alibaba Cloud}}{{Alibaba Cloud}}{2022}]%
        {alibabaclouddatabase}
\bibfield{author}{\bibinfo{person}{{Alibaba Cloud}}.}
  \bibinfo{year}{2022}\natexlab{}.
\newblock \bibinfo{title}{{Alibaba Cloud Databases}}.
\newblock
\newblock
\urldef\tempurl%
\url{https://www.alibabacloud.com/product/databases}
\showURL{%
\tempurl}


\bibitem[\protect\citeauthoryear{{Amazon EC}}{{Amazon EC}}{2015}]%
        {amazon2015amazon}
\bibfield{author}{\bibinfo{person}{{Amazon EC}}.}
  \bibinfo{year}{2015}\natexlab{}.
\newblock \bibinfo{title}{Amazon web services}.
\newblock
\newblock
\urldef\tempurl%
\url{http://aws.amazon.com/es/ec2/}
\showURL{%
\tempurl}


\bibitem[\protect\citeauthoryear{Cao, Feng, Liang, Zhang, Gao, Zhang, and
  Li}{Cao et~al\mbox{.}}{2021}]%
        {cao2021logstore}
\bibfield{author}{\bibinfo{person}{Wei Cao}, \bibinfo{person}{Xiaojie Feng},
  \bibinfo{person}{Boyuan Liang}, \bibinfo{person}{Tianyu Zhang},
  \bibinfo{person}{Yusong Gao}, \bibinfo{person}{Yunyang Zhang}, {and}
  \bibinfo{person}{Feifei Li}.} \bibinfo{year}{2021}\natexlab{}.
\newblock \showarticletitle{LogStore: {A} Cloud-Native and Multi-Tenant Log
  Database}. In \bibinfo{booktitle}{\emph{SIGMOD}}.
  \bibinfo{pages}{2464--2476}.
\newblock


\bibitem[\protect\citeauthoryear{Chandra, Chawda, Kar, Reddy, Shah, and
  Sudarshan}{Chandra et~al\mbox{.}}{2015}]%
        {chandra2015data}
\bibfield{author}{\bibinfo{person}{Bikash Chandra}, \bibinfo{person}{Bhupesh
  Chawda}, \bibinfo{person}{Biplab Kar}, \bibinfo{person}{KV Reddy},
  \bibinfo{person}{Shetal Shah}, {and} \bibinfo{person}{S Sudarshan}.}
  \bibinfo{year}{2015}\natexlab{}.
\newblock \showarticletitle{Data generation for testing and grading SQL
  queries}.
\newblock \bibinfo{journal}{\emph{VLDBJ}} \bibinfo{volume}{24},
  \bibinfo{number}{6} (\bibinfo{year}{2015}), \bibinfo{pages}{731--755}.
\newblock


\bibitem[\protect\citeauthoryear{Chen and Guestrin}{Chen and Guestrin}{2016}]%
        {XGBoost16}
\bibfield{author}{\bibinfo{person}{Tianqi Chen} {and} \bibinfo{person}{Carlos
  Guestrin}.} \bibinfo{year}{2016}\natexlab{}.
\newblock \showarticletitle{{XGBoost}: A Scalable Tree Boosting System}. In
  \bibinfo{booktitle}{\emph{KDD}}. \bibinfo{pages}{785--794}.
\newblock


\bibitem[\protect\citeauthoryear{Conneau, Khandelwal, Goyal, Chaudhary, Wenzek,
  Guzm{\'a}n, Grave, Ott, Zettlemoyer, and Stoyanov}{Conneau
  et~al\mbox{.}}{2020}]%
        {conneau2019unsupervised}
\bibfield{author}{\bibinfo{person}{Alexis Conneau}, \bibinfo{person}{Kartikay
  Khandelwal}, \bibinfo{person}{Naman Goyal}, \bibinfo{person}{Vishrav
  Chaudhary}, \bibinfo{person}{Guillaume Wenzek}, \bibinfo{person}{Francisco
  Guzm{\'a}n}, \bibinfo{person}{{\'E}douard Grave}, \bibinfo{person}{Myle Ott},
  \bibinfo{person}{Luke Zettlemoyer}, {and} \bibinfo{person}{Veselin
  Stoyanov}.} \bibinfo{year}{2020}\natexlab{}.
\newblock \showarticletitle{Unsupervised Cross-lingual Representation Learning
  at Scale}. In \bibinfo{booktitle}{\emph{ACL}}. \bibinfo{pages}{8440--8451}.
\newblock


\bibitem[\protect\citeauthoryear{Copel, Soh, Puca, Manning, and Gollob}{Copel
  et~al\mbox{.}}{2015}]%
        {copeland2015microsoft}
\bibfield{author}{\bibinfo{person}{Marshall Copel}, \bibinfo{person}{Julian
  Soh}, \bibinfo{person}{Anthony Puca}, \bibinfo{person}{Mike Manning}, {and}
  \bibinfo{person}{David Gollob}.} \bibinfo{year}{2015}\natexlab{}.
\newblock \showarticletitle{Microsoft Azure: Planning, Deploying, and Managing
  Your Data Center in the Cloud}.
\newblock \bibinfo{journal}{\emph{Apress}} (\bibinfo{year}{2015}).
\newblock


\bibitem[\protect\citeauthoryear{Damasio, Corvinelli, Godfrey, Mierzejewski,
  Mihaylov, Szlichta, and Zuzarte}{Damasio et~al\mbox{.}}{2019}]%
        {guided2019}
\bibfield{author}{\bibinfo{person}{Guilherme Damasio}, \bibinfo{person}{Vincent
  Corvinelli}, \bibinfo{person}{Parke Godfrey}, \bibinfo{person}{Piotr
  Mierzejewski}, \bibinfo{person}{Alex Mihaylov}, \bibinfo{person}{Jaroslaw
  Szlichta}, {and} \bibinfo{person}{Calisto Zuzarte}.}
  \bibinfo{year}{2019}\natexlab{}.
\newblock \showarticletitle{Guided automated learning for query workload
  re-optimization}.
\newblock \bibinfo{journal}{\emph{PVLDB}} \bibinfo{volume}{12},
  \bibinfo{number}{12} (\bibinfo{year}{2019}), \bibinfo{pages}{2010--2021}.
\newblock


\bibitem[\protect\citeauthoryear{Das, Grbic, Ilic, Jovandic, Jovanovic,
  Narasayya, Radulovic, Stikic, Xu, and Chaudhuri}{Das et~al\mbox{.}}{2019}]%
        {SIGMOD19Index}
\bibfield{author}{\bibinfo{person}{Sudipto Das}, \bibinfo{person}{Miroslav
  Grbic}, \bibinfo{person}{Igor Ilic}, \bibinfo{person}{Isidora Jovandic},
  \bibinfo{person}{Andrija Jovanovic}, \bibinfo{person}{Vivek~R. Narasayya},
  \bibinfo{person}{Miodrag Radulovic}, \bibinfo{person}{Maja Stikic},
  \bibinfo{person}{Gaoxiang Xu}, {and} \bibinfo{person}{Surajit Chaudhuri}.}
  \bibinfo{year}{2019}\natexlab{}.
\newblock \showarticletitle{Automatically Indexing Millions of Databases in
  Microsoft Azure {SQL} Database}. In \bibinfo{booktitle}{\emph{SIGMOD}}.
  \bibinfo{pages}{666--679}.
\newblock


\bibitem[\protect\citeauthoryear{Deep, Gruenheid, Koutris, Naughton, and
  Viglas}{Deep et~al\mbox{.}}{2020}]%
        {deep2020comprehensive}
\bibfield{author}{\bibinfo{person}{Shaleen Deep}, \bibinfo{person}{Anja
  Gruenheid}, \bibinfo{person}{Paraschos Koutris}, \bibinfo{person}{Jeffrey
  Naughton}, {and} \bibinfo{person}{Stratis Viglas}.}
  \bibinfo{year}{2020}\natexlab{}.
\newblock \showarticletitle{Comprehensive and efficient workload compression}.
\newblock \bibinfo{journal}{\emph{PVLDB}} \bibinfo{volume}{14},
  \bibinfo{number}{3} (\bibinfo{year}{2020}), \bibinfo{pages}{418--430}.
\newblock


\bibitem[\protect\citeauthoryear{Duan, Thummala, and Babu}{Duan
  et~al\mbox{.}}{2009}]%
        {VLDB09iTuned}
\bibfield{author}{\bibinfo{person}{Songyun Duan}, \bibinfo{person}{Vamsidhar
  Thummala}, {and} \bibinfo{person}{Shivnath Babu}.}
  \bibinfo{year}{2009}\natexlab{}.
\newblock \showarticletitle{Tuning Database Configuration Parameters with
  iTuned}.
\newblock \bibinfo{journal}{\emph{PVLDB}} \bibinfo{volume}{2},
  \bibinfo{number}{1} (\bibinfo{year}{2009}), \bibinfo{pages}{1246--1257}.
\newblock


\bibitem[\protect\citeauthoryear{Eslami, Tu, Charkhgard, Xu, and Liu}{Eslami
  et~al\mbox{.}}{2019}]%
        {psidb2019}
\bibfield{author}{\bibinfo{person}{Mehrad Eslami}, \bibinfo{person}{Yicheng
  Tu}, \bibinfo{person}{Hadi Charkhgard}, \bibinfo{person}{Zichen Xu}, {and}
  \bibinfo{person}{Jiacheng Liu}.} \bibinfo{year}{2019}\natexlab{}.
\newblock \showarticletitle{PsiDB: A framework for batched query processing and
  optimization}. In \bibinfo{booktitle}{\emph{IEEE BigData}}.
  \bibinfo{pages}{6046--6048}.
\newblock


\bibitem[\protect\citeauthoryear{Gao, Liu, Wu, Li, Wang, and Chen}{Gao
  et~al\mbox{.}}{2022}]%
        {ClusterEA}
\bibfield{author}{\bibinfo{person}{Yunjun Gao}, \bibinfo{person}{Xiaoze Liu},
  \bibinfo{person}{Junyang Wu}, \bibinfo{person}{Tianyi Li},
  \bibinfo{person}{Pengfei Wang}, {and} \bibinfo{person}{Lu Chen}.}
  \bibinfo{year}{2022}\natexlab{}.
\newblock \showarticletitle{ClusterEA: Scalable Entity Alignment with
  Stochastic Training and Normalized Mini-batch Similarities}. In
  \bibinfo{booktitle}{\emph{KDD}}. \bibinfo{pages}{421--431}.
\newblock


\bibitem[\protect\citeauthoryear{Ge, Liu, Chen, Zheng, and Gao}{Ge
  et~al\mbox{.}}{2021a}]%
        {EASY21}
\bibfield{author}{\bibinfo{person}{Congcong Ge}, \bibinfo{person}{Xiaoze Liu},
  \bibinfo{person}{Lu Chen}, \bibinfo{person}{Baihua Zheng}, {and}
  \bibinfo{person}{Yunjun Gao}.} \bibinfo{year}{2021}\natexlab{a}.
\newblock \showarticletitle{Make It Easy: An Effective End-to-End Entity
  Alignment Framework}. In \bibinfo{booktitle}{\emph{SIGIR}}.
  \bibinfo{pages}{777--786}.
\newblock


\bibitem[\protect\citeauthoryear{Ge, Liu, Chen, Zheng, and Gao}{Ge
  et~al\mbox{.}}{2022}]%
        {LargeEA22}
\bibfield{author}{\bibinfo{person}{Congcong Ge}, \bibinfo{person}{Xiaoze Liu},
  \bibinfo{person}{Lu Chen}, \bibinfo{person}{Baihua Zheng}, {and}
  \bibinfo{person}{Yunjun Gao}.} \bibinfo{year}{2022}\natexlab{}.
\newblock \showarticletitle{LargeEA: Aligning Entities for Large-scale
  Knowledge Graphs}.
\newblock \bibinfo{journal}{\emph{{PVLDB}}} \bibinfo{volume}{15},
  \bibinfo{number}{2} (\bibinfo{year}{2022}), \bibinfo{pages}{237--245}.
\newblock


\bibitem[\protect\citeauthoryear{Ge, Wang, Chen, Liu, Zheng, and Gao}{Ge
  et~al\mbox{.}}{2021b}]%
        {ge2021collaborem}
\bibfield{author}{\bibinfo{person}{Congcong Ge}, \bibinfo{person}{Pengfei
  Wang}, \bibinfo{person}{Lu Chen}, \bibinfo{person}{Xiaoze Liu},
  \bibinfo{person}{Baihua Zheng}, {and} \bibinfo{person}{Yunjun Gao}.}
  \bibinfo{year}{2021}\natexlab{b}.
\newblock \showarticletitle{CollaborEM: A Self-supervised Entity Matching
  Framework Using Multi-features Collaboration}.
\newblock \bibinfo{journal}{\emph{TKDE}} (\bibinfo{year}{2021}).
\newblock


\bibitem[\protect\citeauthoryear{Giannella, Han, Pei, Yan, and Yu}{Giannella
  et~al\mbox{.}}{2003}]%
        {giannella2003mining}
\bibfield{author}{\bibinfo{person}{Chris Giannella}, \bibinfo{person}{Jiawei
  Han}, \bibinfo{person}{Jian Pei}, \bibinfo{person}{Xifeng Yan}, {and}
  \bibinfo{person}{Philip~S Yu}.} \bibinfo{year}{2003}\natexlab{}.
\newblock \showarticletitle{Mining frequent patterns in data streams at
  multiple time granularities}.
\newblock \bibinfo{journal}{\emph{Next generation data mining}}
  \bibinfo{volume}{212} (\bibinfo{year}{2003}), \bibinfo{pages}{191--212}.
\newblock


\bibitem[\protect\citeauthoryear{Giannikis, Makreshanski, Alonso, and
  Kossmann}{Giannikis et~al\mbox{.}}{2013}]%
        {sharedb2013}
\bibfield{author}{\bibinfo{person}{Georgios Giannikis}, \bibinfo{person}{Darko
  Makreshanski}, \bibinfo{person}{Gustavo Alonso}, {and}
  \bibinfo{person}{Donald Kossmann}.} \bibinfo{year}{2013}\natexlab{}.
\newblock \showarticletitle{Workload optimization using shareddb}. In
  \bibinfo{booktitle}{\emph{SIGMOD}}. \bibinfo{pages}{1045--1048}.
\newblock


\bibitem[\protect\citeauthoryear{Giannikis, Makreshanski, Alonso, and
  Kossmann}{Giannikis et~al\mbox{.}}{2014}]%
        {shared2014}
\bibfield{author}{\bibinfo{person}{Georgios Giannikis}, \bibinfo{person}{Darko
  Makreshanski}, \bibinfo{person}{Gustavo Alonso}, {and}
  \bibinfo{person}{Donald Kossmann}.} \bibinfo{year}{2014}\natexlab{}.
\newblock \showarticletitle{Shared workload optimization}.
\newblock \bibinfo{journal}{\emph{PVLDB}} \bibinfo{volume}{7},
  \bibinfo{number}{6} (\bibinfo{year}{2014}), \bibinfo{pages}{429--440}.
\newblock


\bibitem[\protect\citeauthoryear{Gr{\"u}nwald}{Gr{\"u}nwald}{2007}]%
        {mdl2007}
\bibfield{author}{\bibinfo{person}{Peter~D Gr{\"u}nwald}.}
  \bibinfo{year}{2007}\natexlab{}.
\newblock \bibinfo{booktitle}{\emph{The minimum description length principle}}.
\newblock \bibinfo{publisher}{MIT press}.
\newblock


\bibitem[\protect\citeauthoryear{Herodotou and Babu}{Herodotou and
  Babu}{2011}]%
        {VLDB11Profiling}
\bibfield{author}{\bibinfo{person}{Herodotos Herodotou} {and}
  \bibinfo{person}{Shivnath Babu}.} \bibinfo{year}{2011}\natexlab{}.
\newblock \showarticletitle{Profiling, What-if Analysis, and Cost-based
  Optimization of MapReduce Programs}.
\newblock \bibinfo{journal}{\emph{PVLDB}} \bibinfo{volume}{4},
  \bibinfo{number}{11} (\bibinfo{year}{2011}), \bibinfo{pages}{1111--1122}.
\newblock


\bibitem[\protect\citeauthoryear{Iyer, Konstas, Cheung, and Zettlemoyer}{Iyer
  et~al\mbox{.}}{2016}]%
        {iyer2016summarizing}
\bibfield{author}{\bibinfo{person}{Srinivasan Iyer}, \bibinfo{person}{Ioannis
  Konstas}, \bibinfo{person}{Alvin Cheung}, {and} \bibinfo{person}{Luke
  Zettlemoyer}.} \bibinfo{year}{2016}\natexlab{}.
\newblock \showarticletitle{Summarizing source code using a neural attention
  model}. In \bibinfo{booktitle}{\emph{ACL}}. \bibinfo{pages}{2073--2083}.
\newblock


\bibitem[\protect\citeauthoryear{Jain, Howe, Yan, and Cruanes}{Jain
  et~al\mbox{.}}{2018}]%
        {query2vec}
\bibfield{author}{\bibinfo{person}{Shrainik Jain}, \bibinfo{person}{Bill Howe},
  \bibinfo{person}{Jiaqi Yan}, {and} \bibinfo{person}{Thierry Cruanes}.}
  \bibinfo{year}{2018}\natexlab{}.
\newblock \showarticletitle{Query2Vec: An Evaluation of NLP Techniques for
  Generalized Workload Analytics}.
\newblock \bibinfo{journal}{\emph{PVLDB}} \bibinfo{volume}{11},
  \bibinfo{number}{5} (\bibinfo{year}{2018}).
\newblock


\bibitem[\protect\citeauthoryear{Jin and Agrawal}{Jin and Agrawal}{2007}]%
        {jin2007frequent}
\bibfield{author}{\bibinfo{person}{Ruoming Jin} {and} \bibinfo{person}{Gagan
  Agrawal}.} \bibinfo{year}{2007}\natexlab{}.
\newblock \showarticletitle{Frequent pattern mining in data streams}.
\newblock In \bibinfo{booktitle}{\emph{Data Streams}}.
  \bibinfo{publisher}{Springer}, \bibinfo{pages}{61--84}.
\newblock


\bibitem[\protect\citeauthoryear{Kennedy, Ajay, Challen, and Ziarek}{Kennedy
  et~al\mbox{.}}{2015}]%
        {kennedy2015pocket}
\bibfield{author}{\bibinfo{person}{Oliver Kennedy}, \bibinfo{person}{Jerry
  Ajay}, \bibinfo{person}{Geoffrey Challen}, {and} \bibinfo{person}{Lukasz
  Ziarek}.} \bibinfo{year}{2015}\natexlab{}.
\newblock \showarticletitle{Pocket data: The need for TPC-MOBILE}. In
  \bibinfo{booktitle}{\emph{Technology Conference on Performance Evaluation and
  Benchmarking}}. Springer, \bibinfo{pages}{8--25}.
\newblock


\bibitem[\protect\citeauthoryear{Kenton and Toutanova}{Kenton and
  Toutanova}{2019}]%
        {bert}
\bibfield{author}{\bibinfo{person}{Jacob Devlin Ming-Wei~Chang Kenton} {and}
  \bibinfo{person}{Lee~Kristina Toutanova}.} \bibinfo{year}{2019}\natexlab{}.
\newblock \showarticletitle{BERT: Pre-training of Deep Bidirectional
  Transformers for Language Understanding}. In
  \bibinfo{booktitle}{\emph{NAACL}}. \bibinfo{pages}{4171--4186}.
\newblock


\bibitem[\protect\citeauthoryear{Krishnan and Gonzalez}{Krishnan and
  Gonzalez}{2015}]%
        {krishnan2015building}
\bibfield{author}{\bibinfo{person}{S.~P.~T. Krishnan} {and}
  \bibinfo{person}{Jose L~Ugia Gonzalez}.} \bibinfo{year}{2015}\natexlab{}.
\newblock \bibinfo{booktitle}{\emph{Building your next big thing with google
  cloud platform: A guide for developers and enterprise architects}}.
\newblock \bibinfo{publisher}{Springer}.
\newblock


\bibitem[\protect\citeauthoryear{Kul, Luong, Xie, Chandola, Kennedy, and
  Upadhyaya}{Kul et~al\mbox{.}}{2018}]%
        {kul2018similarity}
\bibfield{author}{\bibinfo{person}{Gokhan Kul}, \bibinfo{person}{Duc Thanh~Anh
  Luong}, \bibinfo{person}{Ting Xie}, \bibinfo{person}{Varun Chandola},
  \bibinfo{person}{Oliver Kennedy}, {and} \bibinfo{person}{Shambhu Upadhyaya}.}
  \bibinfo{year}{2018}\natexlab{}.
\newblock \showarticletitle{Similarity metrics for SQL query clustering}.
\newblock \bibinfo{journal}{\emph{TKDE}} \bibinfo{volume}{30},
  \bibinfo{number}{12} (\bibinfo{year}{2018}), \bibinfo{pages}{2408--2420}.
\newblock


\bibitem[\protect\citeauthoryear{Le and Mikolov}{Le and Mikolov}{2014}]%
        {doc2vec}
\bibfield{author}{\bibinfo{person}{Quoc Le} {and} \bibinfo{person}{Tomas
  Mikolov}.} \bibinfo{year}{2014}\natexlab{}.
\newblock \showarticletitle{Distributed representations of sentences and
  documents}. In \bibinfo{booktitle}{\emph{International conference on machine
  learning}}. PMLR, \bibinfo{pages}{1188--1196}.
\newblock


\bibitem[\protect\citeauthoryear{Li, Zhou, Sun, Yu, Han, Jin, Li, Wang, and
  Li}{Li et~al\mbox{.}}{2021b}]%
        {li2021opengauss}
\bibfield{author}{\bibinfo{person}{Guoliang Li}, \bibinfo{person}{Xuanhe Zhou},
  \bibinfo{person}{Ji Sun}, \bibinfo{person}{Xiang Yu}, \bibinfo{person}{Yue
  Han}, \bibinfo{person}{Lianyuan Jin}, \bibinfo{person}{Wenbo Li},
  \bibinfo{person}{Tianqing Wang}, {and} \bibinfo{person}{Shifu Li}.}
  \bibinfo{year}{2021}\natexlab{b}.
\newblock \showarticletitle{openGauss: An Autonomous Database System}.
\newblock \bibinfo{journal}{\emph{PVLDB}} \bibinfo{volume}{14},
  \bibinfo{number}{12} (\bibinfo{year}{2021}), \bibinfo{pages}{3028--3041}.
\newblock


\bibitem[\protect\citeauthoryear{Li, Chen, Jensen, and Pedersen}{Li
  et~al\mbox{.}}{2021a}]%
        {li2021trace}
\bibfield{author}{\bibinfo{person}{Tianyi Li}, \bibinfo{person}{Lu Chen},
  \bibinfo{person}{Christian~S Jensen}, {and} \bibinfo{person}{Torben~Bach
  Pedersen}.} \bibinfo{year}{2021}\natexlab{a}.
\newblock \showarticletitle{TRACE: Real-time compression of streaming
  trajectories in road networks}.
\newblock \bibinfo{journal}{\emph{PVLDB}} \bibinfo{volume}{14},
  \bibinfo{number}{7} (\bibinfo{year}{2021}), \bibinfo{pages}{1175--1187}.
\newblock


\bibitem[\protect\citeauthoryear{Li, Huang, Chen, Jensen, and Pedersen}{Li
  et~al\mbox{.}}{2020}]%
        {li2020compression}
\bibfield{author}{\bibinfo{person}{Tianyi Li}, \bibinfo{person}{Ruikai Huang},
  \bibinfo{person}{Lu Chen}, \bibinfo{person}{Christian~S Jensen}, {and}
  \bibinfo{person}{Torben~Bach Pedersen}.} \bibinfo{year}{2020}\natexlab{}.
\newblock \showarticletitle{Compression of uncertain trajectories in road
  networks}.
\newblock \bibinfo{journal}{\emph{PVLDB}} \bibinfo{volume}{13},
  \bibinfo{number}{7} (\bibinfo{year}{2020}), \bibinfo{pages}{1050--1063}.
\newblock


\bibitem[\protect\citeauthoryear{Liu, Wu, Li, Chen, and Gao}{Liu
  et~al\mbox{.}}{2023}]%
        {DualMatch}
\bibfield{author}{\bibinfo{person}{Xiaoze Liu}, \bibinfo{person}{Junyang Wu},
  \bibinfo{person}{Tianyi Li}, \bibinfo{person}{Lu Chen}, {and}
  \bibinfo{person}{Yunjun Gao}.} \bibinfo{year}{2023}\natexlab{}.
\newblock \showarticletitle{Unsupervised Entity Alignment for Temporal
  Knowledge Graphs}. In \bibinfo{booktitle}{\emph{WWW}}.
  \bibinfo{pages}{2528–2538}.
\newblock


\bibitem[\protect\citeauthoryear{Liu, Yin, Zhao, Ge, Chen, Gao, Li, Wang,
  Liang, Tan, and Li}{Liu et~al\mbox{.}}{2022}]%
        {PinSQL}
\bibfield{author}{\bibinfo{person}{Xiaoze Liu}, \bibinfo{person}{Zheng Yin},
  \bibinfo{person}{Chao Zhao}, \bibinfo{person}{Congcong Ge},
  \bibinfo{person}{Lu Chen}, \bibinfo{person}{Yunjun Gao},
  \bibinfo{person}{Dimeng Li}, \bibinfo{person}{Ziting Wang},
  \bibinfo{person}{Gaozhong Liang}, \bibinfo{person}{Jian Tan}, {and}
  \bibinfo{person}{Feifei Li}.} \bibinfo{year}{2022}\natexlab{}.
\newblock \showarticletitle{PinSQL: Pinpoint Root Cause SQLs to Resolve
  Performance Issues in Cloud Databases}. In \bibinfo{booktitle}{\emph{ICDE}}.
  \bibinfo{pages}{2549--2561}.
\newblock


\bibitem[\protect\citeauthoryear{Liu, Ott, Goyal, Du, Joshi, Chen, Levy, Lewis,
  Zettlemoyer, and Stoyanov}{Liu et~al\mbox{.}}{2019}]%
        {liu2019roberta}
\bibfield{author}{\bibinfo{person}{Yinhan Liu}, \bibinfo{person}{Myle Ott},
  \bibinfo{person}{Naman Goyal}, \bibinfo{person}{Jingfei Du},
  \bibinfo{person}{Mandar Joshi}, \bibinfo{person}{Danqi Chen},
  \bibinfo{person}{Omer Levy}, \bibinfo{person}{Mike Lewis},
  \bibinfo{person}{Luke Zettlemoyer}, {and} \bibinfo{person}{Veselin
  Stoyanov}.} \bibinfo{year}{2019}\natexlab{}.
\newblock \showarticletitle{Roberta: A robustly optimized bert pretraining
  approach}.
\newblock \bibinfo{journal}{\emph{arXiv preprint arXiv:1907.11692}}
  (\bibinfo{year}{2019}).
\newblock


\bibitem[\protect\citeauthoryear{Liu, Zhang, Wang, Hou, Yuan, Tian, Zhang, Shi,
  Fan, and He}{Liu et~al\mbox{.}}{2021}]%
        {transformers}
\bibfield{author}{\bibinfo{person}{Yang Liu}, \bibinfo{person}{Yao Zhang},
  \bibinfo{person}{Yixin Wang}, \bibinfo{person}{Feng Hou},
  \bibinfo{person}{Jin Yuan}, \bibinfo{person}{Jiang Tian},
  \bibinfo{person}{Yang Zhang}, \bibinfo{person}{Zhongchao Shi},
  \bibinfo{person}{Jianping Fan}, {and} \bibinfo{person}{Zhiqiang He}.}
  \bibinfo{year}{2021}\natexlab{}.
\newblock \showarticletitle{A Survey of Visual Transformers}.
\newblock \bibinfo{journal}{\emph{CoRR}}  \bibinfo{volume}{abs/2111.06091}
  (\bibinfo{year}{2021}).
\newblock


\bibitem[\protect\citeauthoryear{Ma, Aken, Hefny, Mezerhane, Pavlo, and
  Gordon}{Ma et~al\mbox{.}}{2018}]%
        {ma2018query}
\bibfield{author}{\bibinfo{person}{Lin Ma}, \bibinfo{person}{Dana~Van Aken},
  \bibinfo{person}{Ahmed Hefny}, \bibinfo{person}{Gustavo Mezerhane},
  \bibinfo{person}{Andrew Pavlo}, {and} \bibinfo{person}{Geoffrey~J. Gordon}.}
  \bibinfo{year}{2018}\natexlab{}.
\newblock \showarticletitle{Query-based Workload Forecasting for Self-Driving
  Database Management Systems}. In \bibinfo{booktitle}{\emph{SIGMOD}}.
  \bibinfo{pages}{631--645}.
\newblock


\bibitem[\protect\citeauthoryear{Ma, Yin, Zhang, Wang, Zheng, Jiang, Hu, Luo,
  Li, Qiu, Li, Chen, and Pei}{Ma et~al\mbox{.}}{2020}]%
        {ma2020diagnosing}
\bibfield{author}{\bibinfo{person}{Minghua Ma}, \bibinfo{person}{Zheng Yin},
  \bibinfo{person}{Shenglin Zhang}, \bibinfo{person}{Sheng Wang},
  \bibinfo{person}{Christopher Zheng}, \bibinfo{person}{Xinhao Jiang},
  \bibinfo{person}{Hanwen Hu}, \bibinfo{person}{Cheng Luo},
  \bibinfo{person}{Yilin Li}, \bibinfo{person}{Nengjun Qiu},
  \bibinfo{person}{Feifei Li}, \bibinfo{person}{Changcheng Chen}, {and}
  \bibinfo{person}{Dan Pei}.} \bibinfo{year}{2020}\natexlab{}.
\newblock \showarticletitle{Diagnosing Root Causes of Intermittent Slow Queries
  in Large-Scale Cloud Databases}.
\newblock \bibinfo{journal}{\emph{PVLDB}} \bibinfo{volume}{13},
  \bibinfo{number}{8} (\bibinfo{year}{2020}), \bibinfo{pages}{1176--1189}.
\newblock


\bibitem[\protect\citeauthoryear{Marcus, Negi, Mao, Tatbul, Alizadeh, and
  Kraska}{Marcus et~al\mbox{.}}{2021}]%
        {bao}
\bibfield{author}{\bibinfo{person}{Ryan Marcus}, \bibinfo{person}{Parimarjan
  Negi}, \bibinfo{person}{Hongzi Mao}, \bibinfo{person}{Nesime Tatbul},
  \bibinfo{person}{Mohammad Alizadeh}, {and} \bibinfo{person}{Tim Kraska}.}
  \bibinfo{year}{2021}\natexlab{}.
\newblock \showarticletitle{Bao: Making Learned Query Optimization Practical}.
  In \bibinfo{booktitle}{\emph{SIGMOD}}. \bibinfo{pages}{1275--1288}.
\newblock


\bibitem[\protect\citeauthoryear{Marcus and Papaemmanouil}{Marcus and
  Papaemmanouil}{2016}]%
        {VLDB16WiseDB}
\bibfield{author}{\bibinfo{person}{Ryan Marcus} {and} \bibinfo{person}{Olga
  Papaemmanouil}.} \bibinfo{year}{2016}\natexlab{}.
\newblock \showarticletitle{WiSeDB: {A} Learning-based Workload Management
  Advisor for Cloud Databases}.
\newblock \bibinfo{journal}{\emph{PVLDB}} \bibinfo{volume}{9},
  \bibinfo{number}{10} (\bibinfo{year}{2016}), \bibinfo{pages}{780--791}.
\newblock


\bibitem[\protect\citeauthoryear{Marcus, Negi, Mao, Zhang, Alizadeh, Kraska,
  Papaemmanouil, and Tatbul}{Marcus et~al\mbox{.}}{2019}]%
        {neo}
\bibfield{author}{\bibinfo{person}{Ryan~C. Marcus}, \bibinfo{person}{Parimarjan
  Negi}, \bibinfo{person}{Hongzi Mao}, \bibinfo{person}{Chi Zhang},
  \bibinfo{person}{Mohammad Alizadeh}, \bibinfo{person}{Tim Kraska},
  \bibinfo{person}{Olga Papaemmanouil}, {and} \bibinfo{person}{Nesime Tatbul}.}
  \bibinfo{year}{2019}\natexlab{}.
\newblock \showarticletitle{Neo: {A} Learned Query Optimizer}.
\newblock \bibinfo{journal}{\emph{PVLDB}} \bibinfo{volume}{12},
  \bibinfo{number}{11} (\bibinfo{year}{2019}), \bibinfo{pages}{1705--1718}.
\newblock


\bibitem[\protect\citeauthoryear{Mozafari, Curino, Jindal, and Madden}{Mozafari
  et~al\mbox{.}}{2013}]%
        {mozafari2013performance}
\bibfield{author}{\bibinfo{person}{Barzan Mozafari}, \bibinfo{person}{Carlo
  Curino}, \bibinfo{person}{Alekh Jindal}, {and} \bibinfo{person}{Samuel
  Madden}.} \bibinfo{year}{2013}\natexlab{}.
\newblock \showarticletitle{Performance and resource modeling in
  highly-concurrent OLTP workloads}. In \bibinfo{booktitle}{\emph{SIGMOD}}.
  \bibinfo{pages}{301--312}.
\newblock


\bibitem[\protect\citeauthoryear{{MySQL 8.0 Reference Manual}}{{MySQL 8.0
  Reference Manual}}{2022}]%
        {mysqldigest}
\bibfield{author}{\bibinfo{person}{{MySQL 8.0 Reference Manual}}.}
  \bibinfo{year}{2022}\natexlab{}.
\newblock \bibinfo{title}{{Performance Schema Statement Digests and Sampling}}.
\newblock
\newblock
\urldef\tempurl%
\url{https://dev.mysql.com/doc/refman/8.0/en/performance-schema-statement-digests.html}
\showURL{%
\tempurl}


\bibitem[\protect\citeauthoryear{Nadareishvili, Mitra, McLarty, and
  Amundsen}{Nadareishvili et~al\mbox{.}}{2016}]%
        {nadareishvili2016microservice}
\bibfield{author}{\bibinfo{person}{Irakli Nadareishvili},
  \bibinfo{person}{Ronnie Mitra}, \bibinfo{person}{Matt McLarty}, {and}
  \bibinfo{person}{Mike Amundsen}.} \bibinfo{year}{2016}\natexlab{}.
\newblock \bibinfo{booktitle}{\emph{Microservice architecture: aligning
  principles, practices, and culture}}.
\newblock \bibinfo{publisher}{" O'Reilly Media, Inc."}.
\newblock


\bibitem[\protect\citeauthoryear{Paul, Cao, Li, and Srikumar}{Paul
  et~al\mbox{.}}{2021}]%
        {paul2021database}
\bibfield{author}{\bibinfo{person}{Debjyoti Paul}, \bibinfo{person}{Jie Cao},
  \bibinfo{person}{Feifei Li}, {and} \bibinfo{person}{Vivek Srikumar}.}
  \bibinfo{year}{2021}\natexlab{}.
\newblock \showarticletitle{Database workload characterization with query plan
  encoders}.
\newblock \bibinfo{journal}{\emph{PVLDB}} \bibinfo{volume}{15},
  \bibinfo{number}{4} (\bibinfo{year}{2021}), \bibinfo{pages}{923--935}.
\newblock


\bibitem[\protect\citeauthoryear{Pavlo, Angulo, Arulraj, Lin, Lin, Ma, Menon,
  Mowry, Perron, Quah, Santurkar, Tomasic, Toor, Aken, Wang, Wu, Xian, and
  Zhang}{Pavlo et~al\mbox{.}}{2017}]%
        {CIDR17SelfDriving}
\bibfield{author}{\bibinfo{person}{Andrew Pavlo}, \bibinfo{person}{Gustavo
  Angulo}, \bibinfo{person}{Joy Arulraj}, \bibinfo{person}{Haibin Lin},
  \bibinfo{person}{Jiexi Lin}, \bibinfo{person}{Lin Ma},
  \bibinfo{person}{Prashanth Menon}, \bibinfo{person}{Todd~C. Mowry},
  \bibinfo{person}{Matthew Perron}, \bibinfo{person}{Ian Quah},
  \bibinfo{person}{Siddharth Santurkar}, \bibinfo{person}{Anthony Tomasic},
  \bibinfo{person}{Skye Toor}, \bibinfo{person}{Dana~Van Aken},
  \bibinfo{person}{Ziqi Wang}, \bibinfo{person}{Yingjun Wu},
  \bibinfo{person}{Ran Xian}, {and} \bibinfo{person}{Tieying Zhang}.}
  \bibinfo{year}{2017}\natexlab{}.
\newblock \showarticletitle{Self-Driving Database Management Systems}. In
  \bibinfo{booktitle}{\emph{{CIDR}}}.
\newblock


\bibitem[\protect\citeauthoryear{Psallidas, Agrawal, Sugunan, Ibrahim,
  Karanasos, Camacho-Rodr{\'\i}guez, Floratou, Curino, and
  Ramakrishnan}{Psallidas et~al\mbox{.}}{2022}]%
        {oneprovenance2022}
\bibfield{author}{\bibinfo{person}{Fotis Psallidas}, \bibinfo{person}{Ashvin
  Agrawal}, \bibinfo{person}{Chandru Sugunan}, \bibinfo{person}{Khaled
  Ibrahim}, \bibinfo{person}{Konstantinos Karanasos},
  \bibinfo{person}{Jes{\'u}s Camacho-Rodr{\'\i}guez}, \bibinfo{person}{Avrilia
  Floratou}, \bibinfo{person}{Carlo Curino}, {and} \bibinfo{person}{Raghu
  Ramakrishnan}.} \bibinfo{year}{2022}\natexlab{}.
\newblock \showarticletitle{OneProvenance: Efficient Extraction of Dynamic
  Coarse-Grained Provenance from Database Logs}.
\newblock \bibinfo{journal}{\emph{arXiv preprint arXiv:2210.14047}}
  (\bibinfo{year}{2022}).
\newblock


\bibitem[\protect\citeauthoryear{Reimers and Gurevych}{Reimers and
  Gurevych}{2019}]%
        {sbert}
\bibfield{author}{\bibinfo{person}{Nils Reimers} {and} \bibinfo{person}{Iryna
  Gurevych}.} \bibinfo{year}{2019}\natexlab{}.
\newblock \showarticletitle{Sentence-BERT: Sentence Embeddings using Siamese
  BERT-Networks}. In \bibinfo{booktitle}{\emph{EMNLP}}.
  \bibinfo{pages}{3980--3990}.
\newblock


\bibitem[\protect\citeauthoryear{Richardson and Ruby}{Richardson and
  Ruby}{2008}]%
        {richardson2008restful}
\bibfield{author}{\bibinfo{person}{Leonard Richardson} {and}
  \bibinfo{person}{Sam Ruby}.} \bibinfo{year}{2008}\natexlab{}.
\newblock \bibinfo{booktitle}{\emph{RESTful web services}}.
\newblock \bibinfo{publisher}{" O'Reilly Media, Inc."}.
\newblock


\bibitem[\protect\citeauthoryear{Tang, Wu, Song, Ying, Li, and Chen}{Tang
  et~al\mbox{.}}{2022}]%
        {PreQR}
\bibfield{author}{\bibinfo{person}{Xiu Tang}, \bibinfo{person}{Sai Wu},
  \bibinfo{person}{Mingli Song}, \bibinfo{person}{Shanshan Ying},
  \bibinfo{person}{Feifei Li}, {and} \bibinfo{person}{Gang Chen}.}
  \bibinfo{year}{2022}\natexlab{}.
\newblock \showarticletitle{PreQR: Pre-training Representation for {SQL}
  Understanding}. In \bibinfo{booktitle}{\emph{SIGMOD}}.
  \bibinfo{pages}{204--216}.
\newblock


\bibitem[\protect\citeauthoryear{Theriault and Heney}{Theriault and
  Heney}{1998}]%
        {theriault1998oracle}
\bibfield{author}{\bibinfo{person}{Marlene Theriault} {and}
  \bibinfo{person}{William Heney}.} \bibinfo{year}{1998}\natexlab{}.
\newblock \bibinfo{booktitle}{\emph{Oracle Security}}.
\newblock \bibinfo{publisher}{O'Reilly \& Associates, Inc.}
\newblock


\bibitem[\protect\citeauthoryear{Tran, Morfonios, and Polyzotis}{Tran
  et~al\mbox{.}}{2015}]%
        {tran2015oracle}
\bibfield{author}{\bibinfo{person}{Quoc~Trung Tran},
  \bibinfo{person}{Konstantinos Morfonios}, {and} \bibinfo{person}{Neoklis
  Polyzotis}.} \bibinfo{year}{2015}\natexlab{}.
\newblock \showarticletitle{Oracle Workload Intelligence}. In
  \bibinfo{booktitle}{\emph{SIGMOD}}. \bibinfo{pages}{1669--1681}.
\newblock


\bibitem[\protect\citeauthoryear{Verbitski, Gupta, Saha, Brahmadesam, Gupta,
  Mittal, Krishnamurthy, Maurice, Kharatishvili, and Bao}{Verbitski
  et~al\mbox{.}}{2017}]%
        {amazonuserguide}
\bibfield{author}{\bibinfo{person}{Alexandre Verbitski},
  \bibinfo{person}{Anurag Gupta}, \bibinfo{person}{Debanjan Saha},
  \bibinfo{person}{Murali Brahmadesam}, \bibinfo{person}{Kamal Gupta},
  \bibinfo{person}{Raman Mittal}, \bibinfo{person}{Sailesh Krishnamurthy},
  \bibinfo{person}{Sandor Maurice}, \bibinfo{person}{Tengiz Kharatishvili},
  {and} \bibinfo{person}{Xiaofeng Bao}.} \bibinfo{year}{2017}\natexlab{}.
\newblock \showarticletitle{Amazon aurora: Design considerations for high
  throughput cloud-native relational databases}. In
  \bibinfo{booktitle}{\emph{SIGMOD}}. \bibinfo{pages}{1041--1052}.
\newblock


\bibitem[\protect\citeauthoryear{Wu, Li, Chen, Gao, and Wei}{Wu
  et~al\mbox{.}}{2023}]%
        {wu2023sea}
\bibfield{author}{\bibinfo{person}{Junyang Wu}, \bibinfo{person}{Tianyi Li},
  \bibinfo{person}{Lu Chen}, \bibinfo{person}{Yunjun Gao}, {and}
  \bibinfo{person}{Zhiheng Wei}.} \bibinfo{year}{2023}\natexlab{}.
\newblock \showarticletitle{SEA: A Scalable Entity Alignment System}. In
  \bibinfo{booktitle}{\emph{The 46th International ACM SIGIR Conference on
  Research and Development in Information Retrieval}}.
\newblock


\bibitem[\protect\citeauthoryear{Yoon, Niu, and Mozafari}{Yoon
  et~al\mbox{.}}{2016}]%
        {yoon2016dbsherlock}
\bibfield{author}{\bibinfo{person}{Dong~Young Yoon}, \bibinfo{person}{Ning
  Niu}, {and} \bibinfo{person}{Barzan Mozafari}.}
  \bibinfo{year}{2016}\natexlab{}.
\newblock \showarticletitle{DBSherlock: {A} Performance Diagnostic Tool for
  Transactional Databases}. In \bibinfo{booktitle}{\emph{SIGMOD}}.
  \bibinfo{pages}{1599--1614}.
\newblock


\bibitem[\protect\citeauthoryear{Zhou, Li, Chai, and Feng}{Zhou
  et~al\mbox{.}}{2021}]%
        {learnedrewrite}
\bibfield{author}{\bibinfo{person}{Xuanhe Zhou}, \bibinfo{person}{Guoliang Li},
  \bibinfo{person}{Chengliang Chai}, {and} \bibinfo{person}{Jianhua Feng}.}
  \bibinfo{year}{2021}\natexlab{}.
\newblock \showarticletitle{A Learned Query Rewrite System using Monte Carlo
  Tree Search}.
\newblock \bibinfo{journal}{\emph{PVLDB}} \bibinfo{volume}{15},
  \bibinfo{number}{1} (\bibinfo{year}{2021}), \bibinfo{pages}{46--58}.
\newblock


\bibitem[\protect\citeauthoryear{Zhu, Wu, Chai, Pfadler, Ding, Li, and
  Zhou}{Zhu et~al\mbox{.}}{2022}]%
        {learnedqueryoptim}
\bibfield{author}{\bibinfo{person}{Rong Zhu}, \bibinfo{person}{Ziniu Wu},
  \bibinfo{person}{Chengliang Chai}, \bibinfo{person}{Andreas Pfadler},
  \bibinfo{person}{Bolin Ding}, \bibinfo{person}{Guoliang Li}, {and}
  \bibinfo{person}{Jingren Zhou}.} \bibinfo{year}{2022}\natexlab{}.
\newblock \showarticletitle{Learned Query Optimizer: At the Forefront of
  AI-Driven Databases}. In \bibinfo{booktitle}{\emph{EDBT}}.
  \bibinfo{pages}{1--4}.
\newblock


\bibitem[\protect\citeauthoryear{Zhu, Krishnan, Karanasos, Tarte, Power, Modi,
  Kumar, Zhang, Muthyala, Jurgens, et~al\mbox{.}}{Zhu et~al\mbox{.}}{2021}]%
        {zhu2021kea}
\bibfield{author}{\bibinfo{person}{Yiwen Zhu}, \bibinfo{person}{Subru
  Krishnan}, \bibinfo{person}{Konstantinos Karanasos}, \bibinfo{person}{Isha
  Tarte}, \bibinfo{person}{Conor Power}, \bibinfo{person}{Abhishek Modi},
  \bibinfo{person}{Manoj Kumar}, \bibinfo{person}{Deli Zhang},
  \bibinfo{person}{Kartheek Muthyala}, \bibinfo{person}{Nick Jurgens},
  {et~al\mbox{.}}} \bibinfo{year}{2021}\natexlab{}.
\newblock \showarticletitle{Kea: Tuning an exabyte-scale data infrastructure}.
  In \bibinfo{booktitle}{\emph{SIGMOD}}. \bibinfo{pages}{2667--2680}.
\newblock


\bibitem[\protect\citeauthoryear{Zolaktaf, Milani, and Pottinger}{Zolaktaf
  et~al\mbox{.}}{2020}]%
        {facilitating2020}
\bibfield{author}{\bibinfo{person}{Zainab Zolaktaf}, \bibinfo{person}{Mostafa
  Milani}, {and} \bibinfo{person}{Rachel Pottinger}.}
  \bibinfo{year}{2020}\natexlab{}.
\newblock \showarticletitle{Facilitating SQL query composition and analysis}.
  In \bibinfo{booktitle}{\emph{SIGMOD}}. \bibinfo{pages}{209--224}.
\newblock


\end{thebibliography}

\end{document}